\pdfoutput=1

\documentclass[aps,prb,reprint,groupedaddress]{revtex4-1}
\usepackage{geometry} 
\usepackage{hyperref}
\usepackage{natbib}
\usepackage{bm}
\usepackage{amsmath}
\usepackage{amssymb}
\usepackage{mathrsfs}
\usepackage{graphicx}
\usepackage{fullpage}
\usepackage{color}
\usepackage{xcolor}

\begin{document}

\title{Surface Nematic Order in Iron Pnictides}
\author{Kok Wee Song and Alexei E. Koshelev}
\affiliation{Materials Science Division, Argonne National Laboratory, Illinois, 60439, USA}
\date{\today }

\begin{abstract}
Electronic nematicity plays important role in iron-based superconductors.
These materials have layered structure and theoretical description of their magnetic and nematic transitions has been well established in two-dimensional approximation, i.e., when the layers can be treated independently. However, the interaction between iron layers mediated by electron tunneling may cause non-trivial three-dimensional behavior. 
Starting from the simplest model for orbital nematic in a single layer, we investigate the influence of interlayer tunneling on bulk nematic order and possible preemptive state where this order is only formed near the surface.
We found that the interlayer tunneling suppresses the bulk nematicity 
which makes favorable formation of a surface nematic above the bulk transition temperature.
The purely electronic tunneling Hamiltonian, however, favors alternating from layer-to-layer nematic order parameter in the bulk. The uniform bulk state typically observed experimentally may be stabilized by the coupling with the elastic lattice deformation. Depending on strength of this coupling, we found three regimes: (i) surface nematic and alternating bulk order, (ii) surface nematic and uniform bulk order, and (iii) uniform bulk order without the intermediate surface phase. 
The intermediate surface-nematic state may resolve the current controversy about the existence of the weak nematic transition in the compound BaFe$_2$As$_{2-x}$P$_{x}$.
\end{abstract}


\maketitle

\section{Introduction}

Iron-based superconductors are multiple-band layered materials. \cite{Paglione:2010fk,Stewart:2011fk,Hosono2015399} 
The correlations of itinerant electrons in different bands create
several collective excitations: magnetic, orbital, superconducting. 
The complex interplay between these excitations can be tuned by doping or pressure leading to rich phase diagrams with antiferromagnetic, nematic, and superconducting phases.

Parent materials have stripe antiferromagnetic and structural tetragonal-to-orthorhombic phase
transition. The corresponding orders can be either established simultaneously, via a
single first-order transition, or via two sequential second-order phase transitions with
the structural transition always  preceding the antiferromagnetic one. In particular, the
first scenario is realized in Ba$_{1-x}$K$_x$Fe$_2$As$_2$ (122 structure),
\cite{AvciPhysRevB.85.184507} while the second scenario is realized above threshold
doping level in Ba(Fe$_{1-x}$Co$_{x}$)$_{2}$As$_{2}$,
\cite{KimPhysRevB.83.134522,RotunduPhysRevB.84.092501} in ReFeAsO$_{1-x}$F$_x$
compounds (1111 structure), where Re is the rare-earth element (La, Pr, Sm, Ce),
\cite{ZhaoNatMat08,HuangPhysRevB.78.054529,LuetkensNatMat09,RotunduPhysRevB.80.144517,JeschePhysRevB.81.134525} and in Na$_{1-\delta}$FeAs. \cite{ChenPhysRevLett.102.227004,LiPhysRevB.80.020504}

When the four-fold crystal symmetry breaks, at least three types of order emerge
simultaneously: (i) orthorhombic lattice deformation, (ii) spin Ising-nematic order lifting
degeneracy between the stripe-antiferromagnetic fluctuations in two orthogonal directions,
\cite{PhysRevB.77.224509,PhysRevB.78.020501,PhysRevB.85.024534,PhysRevB.91.214515} and (iii)
energy split between the $d_{zx}$ and $d_{zy}$ Fe orbitals leading to density difference between the
two electron bands (ferro-orbital order).\cite{PhysRevB.79.054504,LvPhysRevB.80.224506,LeePhysRevLett.103.267001,ChenPhysRevB.82.100504}
Two latter orders are realizations of electronic nematicity \cite{annurev-conmatphys-070909-103925} 
and, most likely, the orthorhombic deformation is its consequence.
This interpretation is supported by measurement of unusual resistivity anisotropy very sensitive to elastic stress, \cite{Chu13082010,*Chu10082012,PhysRevB.81.184508} softening of the elastic shear modulus, \cite{FernandesPhysRevLett.105.157003,Yoshizawa:JPSJ81.2012,BohmerPhysRevLett.112.047001} optical conductivity, \cite{0295-5075-93-3-37002,*NakajimaPhysRevLett.109.217003}
and asymmetric shifts of orbital energies observed by ARPES. \cite{Yi26042011}
Whether the electronic nematicity has predominantly magnetic or orbital origin is the subject of ongoing debate. \cite{Fernandes:2014sf,chubukov2016magnetism}

In addition to the strong and well-established ``main'' simultaneous AFM and structural transition
in the parent and P-doped 122 materials, torque, \cite{Kasahara:2012fk} NMR,\cite{doi:10.7566/JPSJ.84.043705} and optical-pumping\cite{2015arXiv150703981T} experiments 
suggested existence of an intermediate nematic phase
with broken C4 symmetry emerging at temperatures $\sim$ 20K above the main transition. These observations are also consistent with finite orbital splitting persisting above the bulk transition found for
unstressed Ba(Fe$_{1-x}$Co$_{x}$)$_{2}$As$_{2}$ crystals by ARPES. \cite{Yi26042011} On the other
hand, the recent high-resolution specific heat measurements\cite{PhysRevB.91.094512} clearly excluded
possibility of the bulk phase transition in this temperature range.

One possibility to resolve this controversy is to assume the existence of a preemptive
state, at which the nematic order nucleates first only at the surface and decays inside the bulk. In
this paper, we investigate the role of interlayer tunneling on bulk nematic order and
possibility of such preemptive surface nematic. We use the simplest model for a single
layer in which we take into account only electron pockets and the orbital order appears
due to Pomeranchuk instability caused by interaction between the pockets. We found that
the interlayer tunneling suppresses the bulk transition. As a consequence, it is favorable
for the nematic order to form near the surface first. We found that the purely electronic
tunneling Hamiltonian favors bulk nematic order parameter which alternates from
layer to layer. Such alternating state is not realized in iron pnictides. 
The uniform bulk state observed experimentally may be stabilized by the coupling with the elastic lattice
deformation. 
The similar nematic-lattice coupling has also been studied in Ref.\ \onlinecite{PhysRevLett.111.047004} for the single-layer model using Monte-Carlo simulations and it was demonstrated that this coupling lifts the nematic transition above the magnetic transition. In this paper, depending on strength of the nematic-lattice coupling, we found three regimes: (i) surface
nematic and alternating bulk order, (ii) surface nematic and uniform bulk order, and (iii)
uniform bulk order without intermediate surface phase. 
In the first two scenarios, the nematic order at the onset of ordering instability has strong spatial variation in the out-of-plane direction with maximum at the surface. In the later discussions, this type of instability will be referred to as surface instability. 
	\footnote{
We note that in the case of continuous transition we consider, the surface instability actually smears the bulk transition in finite-size samples. With decreasing temperature the nematic order smoothly extends over larger distances away from the surface. In macroscopic samples, however, the bulk transition becomes a very sharp crossover occurring at the bulk transition temperature, the larger sample the sharper crossover.  In contrast, in the case of first-order bulk transition, not considered in this paper,  there should be two distinct phase transitions.
		}
The decay length typically is of the order of several layer spacings from the surface. In the vicinity of transition between the second and third regimes, the decay length rapidly increases, and eventually diverges at the transition to the third regime. 
In the third regime the order parameter nucleates mostly uniformly inside 
the sample with some suppression near the surface (bulk instability).

The paper is organized as follows. In section \ref{model}, we discuss the model of the nematic order
for finite-size system. In section \ref{TS}, we consider the system free energy, locate the nematic instability,
and  find the stable ground state at the onset of transition for both infinite and finite-size
system. In section \ref{distortion}, we discuss effects of the lattice on the electronic nematic order.
We summarize and conclude the paper in section \ref{finale}.



\section{Model}\label{model}

\subsection{Single layer}\label{model:mono}

We start from a single-layer model Hamiltonian
\begin{equation}\label{H1}
H=\sum_{\alpha,\mathbf{k}}\varepsilon^\alpha_{\mathbf{k}}c^\dagger_{\alpha,\mathbf{k}s}c_{\alpha,\mathbf{k}s}-\frac{uS}{2}\sum_{\mathbf{q}}\rho_\mathbf{q}\rho_\mathbf{-q},
\end{equation}
where $s$ is the spin (the summation is implicitly assumed), $S$ is the total area of the layer, $\mathbf{k}=(k_x, k_y)$ is the in-plane momentum, $\alpha= X \text{ and } Y$ represent the electron pockets at $(\pi,0)$ and $(0,\pi)$ in the 1-Fe Brillouin zone respectively. The electrons energy dispersions near the $X$ and $Y$ pockets are $\varepsilon^{X}_{\mathbf{k}}=\frac{k^2_x}{2m_x}+\frac{k^2_y}{2m_y}$ and $\varepsilon^{Y}_\mathbf{k}=\frac{k^2_x}{2m_y}+\frac{k^2_y}{2m_x}$ with the band masses $m_x$ and $m_y$, where $\mathbf{k}$ is measured from $(\pi,0)$ in the $X$ pocket and from $(0,\pi)$ in the $Y$ pocket.
$\rho_\mathbf{q}$ is a charge collective mode with
\begin{align*}
 \rho_\mathbf{q}&=\frac{1}{S}\sum_{\mathbf{k}}(c^\dagger_{Y,\mathbf{k}+\mathbf{q},s}c_{Y,\mathbf{k}s}-c^\dagger_{X,\mathbf{k}+\mathbf{q},s}c_{X,\mathbf{k}s}).
\end{align*}
The Hamiltonian is invariant under the exchange between $X$ and $Y$ which preserves the 4-fold rotational symmetry. This Hamiltonian has been discussed in Ref.\ \onlinecite{Yamase:2013aa} and can be obtained from a more general itinerant model.\cite{PhysRevB.78.134512} We do not include the hole bands in the middle of the Brillouin zone which do not play role in the consideration.

If the coupling constant $u$ is large enough and positive, the model can give rise to Pomeranchuk instability at $\mathbf{q}=(0,0)$, and the Fermi surface (FS) is distorted in the ground state. The nematic order parameter of the model \eqref{H1} can be identified as
\begin{equation*}
\Delta_\mathbf{q}=u\langle \rho_{\mathbf{q}}\rangle.
\end{equation*}
where $\langle\dots\rangle=\text{Tr}(\dots e^{-\beta (H-\mu\mathcal{N})})/\text{Tr}e^{-\beta (H-\mu\mathcal{N})}$ is a trace over all many-body quantum states with $\beta=1/T$, chemical potential $\mu$, and total number operator $\mathcal{N}=\sum_{\mathbf{k}\alpha}c^\dagger_{\alpha,\mathbf{k}s}c_{\alpha,\mathbf{k}s}$. This order parameter measures the difference between the electron densities in $X$ and $Y$  pockets in the ground state.  Since the main orbital component of the itinerant electrons at X and Y pockets is $d_{zx}$ and $d_{zy}$ orbital respectively,\cite{1367-2630-11-2-025016} this nematic order has a close connection with the orbital ordering. 

We will use the standard mean-field approximation, see, e.g., Ref. \onlinecite{FradkinQFT}, which assumes that the fluctuations $\langle\delta\rho_\mathbf{q}\delta\rho_{-\mathbf{q}}\rangle$ with $\delta\rho_\mathbf{q}=\rho_\mathbf{q}-\langle\rho_\mathbf{q}\rangle$ are small, and  $\delta\rho_\mathbf{q}\delta\rho_{-\mathbf{q}}$ can be neglected in Eq. \eqref{H1}. This yields the mean-field Hamiltonian
\begin{equation*}
H\simeq\sum_{\alpha,\mathbf{k}}\varepsilon^\alpha_{\mathbf{k}}c^\dagger_{\alpha,\mathbf{k},s}c_{\alpha,\mathbf{k}s}-S\sum_{\mathbf{q}}\rho_\mathbf{q}\Delta_{-\mathbf{q}}+\sum_\mathbf{q}\frac{S|\Delta_{\mathbf{q}}|^2}{2u}.
\end{equation*}
Furthermore, one may assume that homogeneous order is the most energetically favorable state in an ideal crystal, $\Delta_\mathbf{q}=\Delta\delta(\mathbf{q})$. This immediately leads to 
\begin{equation}\label{HMF1}
H\simeq\sum_{\alpha,\mathbf{k}}(\varepsilon^\alpha_{\mathbf{k}}+V^\alpha)c^\dagger_{\alpha,\mathbf{k},s}c_{\alpha,\mathbf{k}s}+\frac{S\Delta^2}{2u},
\end{equation}
where $V^X=\Delta$ and $V^Y=-\Delta$.

We remark that the particular form of the microscopic model in Eq.\ \eqref{H1} is not crucial for our study. The approach in this paper can also be applied to other microscopic models which have the similar effective mean-field Hamiltonian. Furthermore, the general framework of nematic order induced by Pomeranchuk instabilities was first discussed in the $d$-wave nematic order,\cite{doi:10.1143/JPSJ.69.332, *doi:10.1143/JPSJ.69.2151,PhysRevLett.85.5162} and also was used in Refs. \onlinecite{PhysRevB.72.035114,PhysRevB.68.245109,PhysRevB.72.035114,PhysRevB.70.155110}. Recently, Pomeranchuk instability in FeSC has also been investigated using renormalization group\cite{chubukov2016magnetism}
and quantum Monte Carlo\cite{dumitrescu2015superconductivity} techniques, see also subsequent discussion on comparing the Monte-Carlo and analytic results.\cite{PhysRevB.93.165141}
In the next section we consider tunneling terms for the layered systems.

\subsection{Interlayer tunneling}

In this section, we discuss the effective three-dimensional model which takes into account interlayer electronic tunneling. This model will be used for analyzing  
nematic transition in layered materials. The building block of the multilayer model in this paper is the single-layer mean-field Hamiltonian from Eq.\ \eqref{H1} with homogeneous nematic order within the plane, $\mathbf{q}=(0,0)$. We will start with the simple model taking into account only nearest neighbor interlayer tunneling terms. This model does not mix the X- and Y- pockets and allows for several analytical results in the limit of weak interlayer hopping constant. Unfortunately, this simple model does not quite describe situation for real crystal structure in 122 materials, where interlayer hoppings via pnictogen atoms extend beyond nearest neighbors, break 4-fold rotational symmetry, and mix  X- and Y- pockets.\cite{PhysRevB.86.144519} We, therefore, will extend our interlayer model to include these effects. 

\subsubsection{Nearest-neighbor hopping}

With the single-layer Hamiltonian \eqref{HMF1}, the $N$-layer system with nearest-neighbor interlayer hopping can be modeled by the following mean-field Hamiltonian 
\begin{equation}\label{HN}
\begin{split}
\mathcal{H}_{N}=&\sum^{N}_{\ell=1}\Big[\sum_{\alpha,\mathbf{k}}(\varepsilon^\alpha_{\mathbf{k}}+V^\alpha_{\ell})c^\dagger_{\alpha\ell\mathbf{k}s}c_{\alpha\ell\mathbf{k}s}+\frac{S\Delta^2_{\ell}}{2u}\Big]\\
&-t_z\sum^{N-1}_{\ell=1}\sum_{\alpha,\mathbf{k}}c^\dagger_{\alpha\ell\mathbf{k}s}c_{\alpha,\ell+1,\mathbf{k}s}+h.c.,
\end{split}
\end{equation}
where each layer is labeled by $\ell=1,\dots, N$, and $V^X_\ell=\Delta_\ell$, $V^Y_\ell=-\Delta_\ell$ are the nematic order parameters in the $\ell$-th layer. Here, we let the amplitude of the order parameters vary alone the z-direction, since in a finite-size system translational symmetry is broken explicitly at the surfaces. Since the important electronic correlations are intralayer, we assumed that the direct overlapping between the Fe-orbitals in different layers are negligible and ignore the interlayer electron-electron interaction in the model.


\subsubsection{Hoppings beyond the nearest neighbors: X-Y hybridization}

\begin{figure}[htbp] 
   \centering
   \includegraphics[scale=0.27]{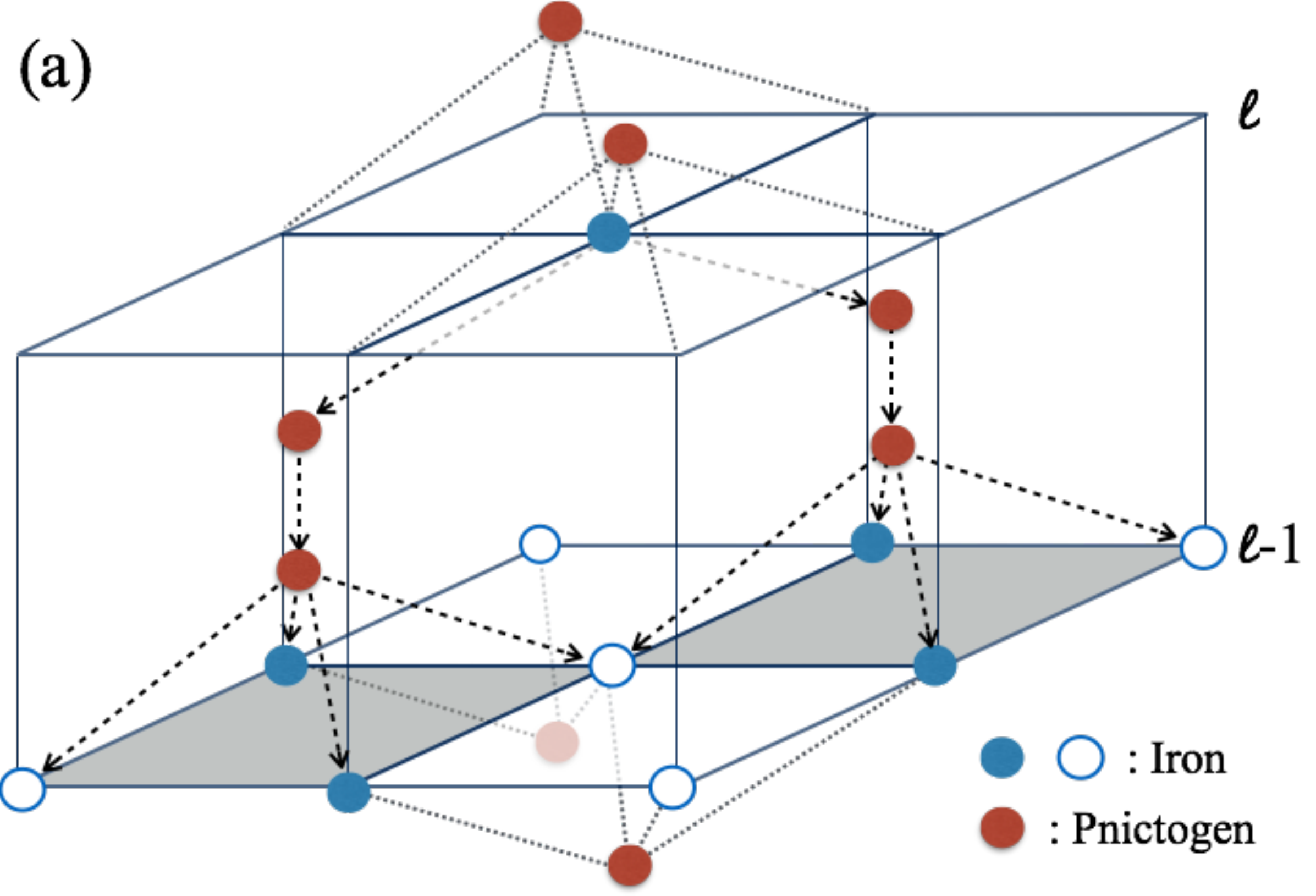}
   \includegraphics[scale=0.26]{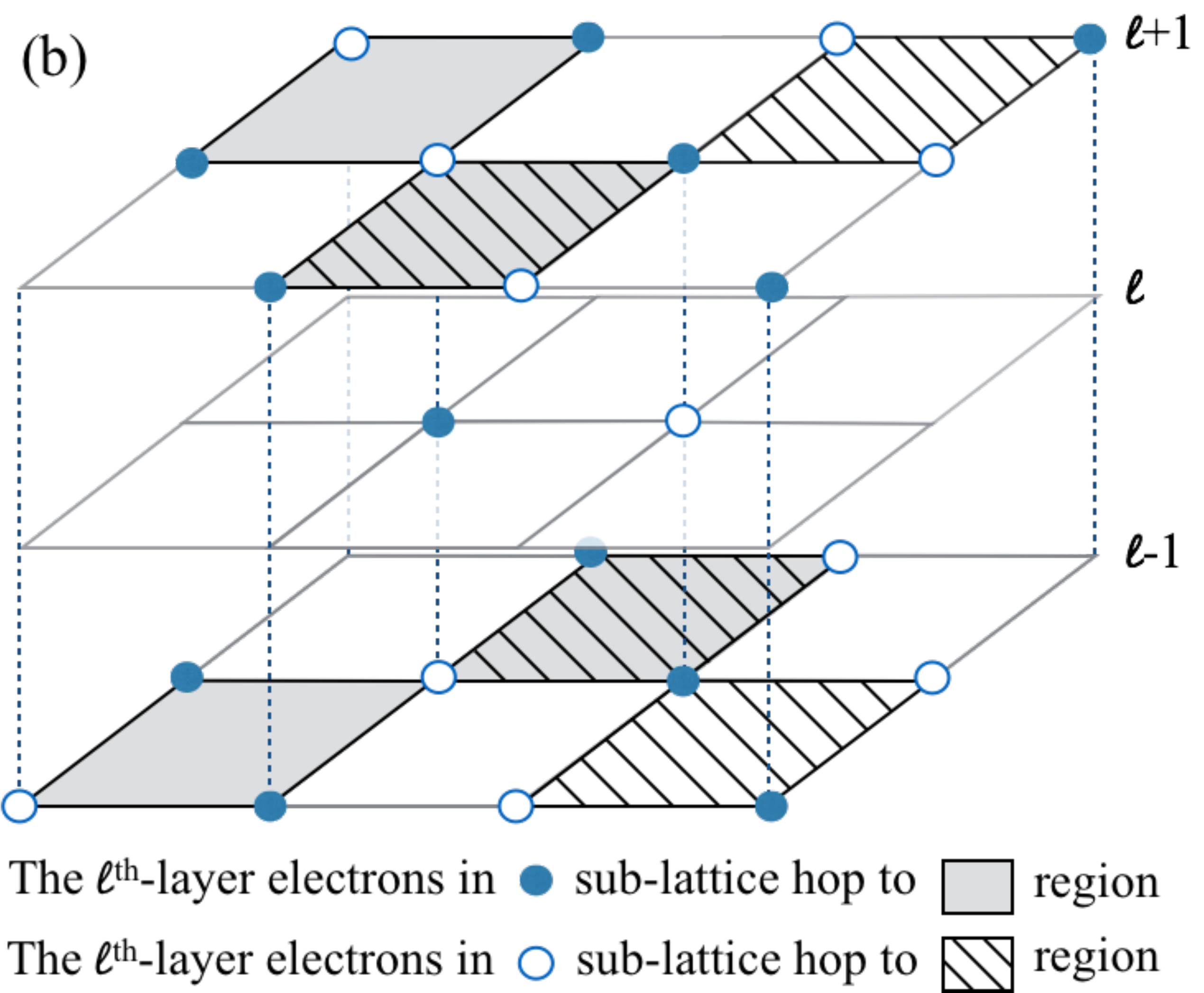}
   \caption{ Interlayer tunneling in 122 crystal: (a) The electron in the upper layer can tunnel to the shaded region in the lower layer via the pnictogens. These hopping processes should be treated at equal footing, since they all have equal tunneling probability. (b) The electron in the $\ell$-th layer odd (even) sub-lattice hopping to the shaded (tilted-lines) region in $(\ell\pm1)$-th layer. The blue open circles and dots are the even and odd sublattice in the iron layer respectively, and the pnictogens in the lattice are not shown in the diagram. }
   \label{fig:FeAs}
\end{figure}

The nearest-neighbor tunneling in Eq.\ \eqref{HN} gives the simplest model for the interlayer coupling.  
However, for the crystalline structure of the 122 family of iron pnictides, such as BaFe$_2$As$_2$,
the interlayer tunneling is a complicated process which involves the hopping between the Fe-orbitals and pnictogen orbitals  (see Fig. \ref{fig:FeAs}a). As a consequence, the interlayer hoppings beyond the nearest neighbor are as important as  the nearest-neighbor hopping.  

This leads to two modifications in Eq.\ \eqref{HN} (see Appendix \ref{app:HXY}). First, due to the hoppings to the second neighbors, the hopping term becomes $\mathbf{k}$-dependent.
\begin{equation}\label{Htun'}
\begin{split}
\mathcal{H}'_{\mathrm{tun}}=-\sum^{N-1}_{\ell=1}\sum_{\mathbf{k}}s_\mathbf{k}
(c^\dagger_{\alpha\ell\mathbf{k}s}c_{\alpha,\ell+1,\mathbf{k}s}+h.c.),
\end{split}
\end{equation}
where $s_\mathbf{k}=2t_z(\cos k_x+\cos k_y )$.
Second, the hoppings to the third neighbors break the 4-fold rotation symmetry and this introduces the hybridization between the electrons in $X$- and $Y$- pockets.\cite{PhysRevB.86.144519} The hybridization Hamiltonian is given by
\begin{equation}\label{Hybpp}
\begin{split}
\mathcal{H}_{\mathrm{hyb}}=\sum_{\ell=1}^{N-1}(-1)^{\ell+1}\sum_{\mathbf{k}}\lambda_\mathbf{k}(c^\dagger_{X \ell\mathbf{k}s}c_{Y,\ell+1,\mathbf{k}s}\\
+c^\dagger_{Y \ell\mathbf{k}s}c_{X,\ell+1,\mathbf{k}s} + h.c.),
\end{split}
\end{equation}
where $\lambda_\mathbf{k}=2t_z'\sin k_x\sin k_y$.


\section{Nematic phase transition}\label{TS}


\subsection{Single-layer nematic transition}
First, we consider the transition temperature for a single-layer system, Eq. \eqref{HMF1}. 
The free energy per unit area is
\begin{equation}\label{F1}
\begin{split}
F_1[\Delta]
&=\frac{\Delta^2}{2u}-2\sum_{\alpha}\int_\mathbf{k}\ln\Big[1+e^{-\beta(\xi^\alpha_{\mathbf{k}}+V^\alpha)}\Big],
\end{split}
\end{equation}
where 
$\xi^\alpha_\mathbf{k}=\varepsilon^\alpha_\mathbf{k}-\mu$, and $\frac{1}{S}\sum_\mathbf{k}\to\int_\mathbf{k}=\int\frac{d^2k}{(2\pi)^2}$. The factor of two in the second term in Eq. \eqref{F1} accounts for the spin degeneracy. To obtain the transition temperature, one can expand the free energy near the critical point and focus on the quadratic order term in $F_1$. 
Namely,
\begin{equation*}
F_1[\Delta]\simeq F_1[0]+\frac{r_1}{2}\Delta^2
\end{equation*}
with the inverse nematic susceptibility
\begin{equation}
r_1=\frac{1}{u}+2\sum_{\alpha}\int_\mathbf{k}n'_F(\xi^\alpha_{\mathbf{k}}),
\label{r1}
\end{equation}
where $n_F(z)=[1+\exp(\beta z)]^{-1}$ is the Fermi-Dirac distribution function.
At the transition temperature, the inverse nematic susceptibility changes sign. Therefore, the nematic phase transition temperature $T_0$ can be determined by setting this coefficient to zero and solve for $\beta$. 

Furthermore, since the electronic modes far from the FS are suppressed by the Fermi-Dirac distribution factor, the upper limit of the momentum integral can be evaluated as $\int_\mathbf{k}=\frac{\tilde{m}}{2\pi}\int^\infty_0d\varepsilon$ with $\varepsilon=k_x^2/(2m_x)+k_y^2/(2m_y)$ and $\tilde{m}=\sqrt{m_xm_y}$. This yields the explicit result for the single-layer inverse susceptibility,
\begin{equation}
r_1=\frac{1}{u}-\frac{\tilde{m}}{\pi}\Big(1+\tanh\frac{\beta\mu}{2}\Big).
\label{r1result}
\end{equation}
In order to have a non-trivial solution, the coupling constant must satisfy the following condition
\begin{equation}\label{constrain}
\frac{1}{4}< \frac{\tilde{m}u}{2\pi}<\frac{1}{2},
\end{equation} 
since $0\leq\tanh\frac{\beta\mu}{2}\leq 1$. For $\frac{\tilde{m}u}{2\pi}\leq1/4$, no nematic order can be sustained in any temperature. For $\frac{\tilde{m}u}{2\pi}\geq 1/2$, the system is in the nematic phase for all temperatures with no phase transition. Therefore, the rest of the paper, we only consider the coupling constant in the region given by Eq. \eqref{constrain}.

\subsection{Bulk nematic transition}\label{Bulk}

In this section, we consider the bulk system in thermodynamic limit, $N\rightarrow\infty$. We start with the simplest model described by the Hamiltonian which takes into account only the nearest-neighbor hopping, Eq.\ \eqref{HN}. Then, we will generalize the model by including the hopping terms  \eqref{Htun'} and \eqref{Hybpp} beyond the nearest neighbor.

\subsubsection{Nearest-neighbor hopping}
\label{Bulk-NNHop}
The free energy of the system  in the $N\to\infty$ limit  is convenient to calculate in the momentum space. 
The Fourier transformation of the field operators are 
\begin{equation*}
c_{\alpha\ell\mathbf{k}s}\!=\!\sum_{\mathbf{k}_z}\frac{e^{-ik_z\ell}}{\sqrt{N}}c_{\alpha k_z\mathbf{k}s},\quad c_{\alpha k_z\mathbf{k}s}\!=\!\sum^\infty_{\ell=-\infty}\frac{e^{ik_z\ell}}{\sqrt{N}}c_{\alpha \ell\mathbf{k}s}.
\end{equation*}
The momentum space representation of the Hamiltonian \eqref{HN} is
\begin{equation}
\mathcal{H}\!=\!
\sum_{\mathbf{k}k'_zk_z}\!\psi^\dagger_{k_z}(\hat{H}_b+\hat{V}_b)\psi_{k'_z}
\!+\!\frac{S}{2u}\frac{1}{N}\sum_q|\tilde{\Delta}_q|^2,
\end{equation}
where $k_z\in[-\pi,\pi]$ is the out-of-plane momentum, $|\tilde{\Delta}_{q}|^2=\tilde{\Delta}_{q}\tilde{\Delta}_{-q}$, and $\psi^\dagger_{k_z}=(c^\dagger_{X k_z \mathbf{k}s}, c^\dagger_{Y k_z \mathbf{k}s})$. The $2\times2$ matrices are
\begin{align}
\hat{H}_b&=\begin{pmatrix}
\varepsilon^X_\mathbf{k}\!-\!2t_z\cos k_z&0\\
0&\varepsilon^Y_\mathbf{k}\!-\!2t_z\cos k_z
\end{pmatrix}\delta_{k_z,k_z'},\\
\hat{V}_b&=\mathscr{V}_{k_zk_z'}=\frac{1}{N}\begin{pmatrix}
\tilde{\Delta}_{k_z-k_z'}&0\\
0&-\tilde{\Delta}_{k_z-k_z'}
\end{pmatrix}.
\end{align}
Furthermore, the order parameter in the momentum space is
\[
\tilde{\Delta}_q=\sum^\infty_{\ell=-\infty} \Delta_{\ell}e^{-iq\ell}, \quad \Delta_\ell=\int_{q}\tilde{\Delta}_q e^{iq\ell},
\]
where we replaced $\frac{1}{N}\sum_q$ by $\int_q=\int\frac{dq}{2\pi}$ in the limit $N\to\infty$. Note that $\tilde{\Delta}_q$ is a periodic function of $q$ with period $2\pi$, $\tilde{\Delta}_q=\tilde{\Delta}_{q\pm2\pi}$, and we choose the range of $q$ to be $[-\pi,\pi]$. The free energy of this simple model can be immediately written down as
\begin{equation*}
\begin{split}
F_b[\tilde{\Delta}_q]\!=&\!-\frac{1}{\beta S} \ln \text{Tr}e^{-\beta(\mathcal{H}-\mu\mathcal{N})}\\
\!=&\!\int_{q}\frac{|\tilde{\Delta}_{q}|^2}{2u}
-\frac{2}{\beta}\frac{1}{S}\text{tr}\ln[i\omega_n\!-\!\hat{H}_b-\hat{V}_b+\mu],
\end{split}
\end{equation*}
where $\omega_n=2\pi T(n+1/2)$ are the fermionic Matsubara frequencies and
the trace is the sum over all $\omega_n$, $\mathbf{k}$, $k_z$, and $\alpha$.
To find the transition temperature, we expand the functional with respect to $\tilde{\Delta}_q$ up to second order,
\begin{equation}\label{Fb}
\begin{split}
F_b[&\tilde{\Delta}_q]\simeq F_b[0]+\int_q\frac{|\tilde{\Delta}_{q}|^2}{2u}\\
&+\frac{1}{\beta}\sum_{\omega_n}\sum_{k_z,k^{\prime}_z}\int_{\mathbf{k}}\text{tr}(\mathscr{G}_{k_z}\mathscr{V}_{k_z,k'_z}\mathscr{G}_{k'_z}\mathscr{V}_{k'_z,k_z})
\end{split}
\end{equation}
with $\mathscr{G}_{k_z}=(i\omega_n-\hat{H}_b+\mu)^{-1}$. Carrying out the trace explicitly in Eq.\ \eqref{Fb}, we obtain
\begin{equation}
F_b[\tilde{\Delta}_q]=F_b[0]+\frac{1}{2}\int_qr_{b,q}|\tilde{\Delta}_q|^2,
\end{equation}
where the inverse nematic susceptibility
 \begin{equation}
 r_{b,q}=\frac{1}{u}-\sum_{\alpha}\int_{\mathbf{k}}\int_{k_z}
 \frac{n_F(z^\alpha_{k_z})-n_F(z^\alpha_{k_z+q})}
 {t_z[\cos k_z-\cos (k_z+q)]}
 \end{equation}
with $\frac{1}{N}\sum_{k_z}\to\int_{k_z}=\int\frac{d k_z}{2\pi}$, $z^\alpha_{k_z}=\xi^\alpha_\mathbf{k}-2t_z\cos k_z$ and $\xi^\alpha_\mathbf{k}=\varepsilon^\alpha_\mathbf{k}-\mu$. In the small-$t_z$ limit, expansion over $t_z$ gives 
\begin{equation}\label{rbqsmalltz}
 r_{b,q}\simeq r_1+\Big(1+\frac{\cos q}{2}\Big)r_t,
\end{equation}
where $r_1$ is given by Eq.\ \eqref{r1result} and $r_t=\frac{8}{3}t_z^2\int_\mathbf{k}n'''_F(\xi^\alpha_\mathbf{k})$.
Performing integration over the in-plane momentum, we obtain 
\begin{equation}
r_t=\frac{2\tilde{m}\beta^2t^2_z}{3\pi}\tanh\Big(\frac{\beta\mu}{2}\Big)\mathrm{sech}^2\Big(\frac{\beta\mu}{2}\Big).
\label{rt}
\end{equation}
Similar to the single-layer case, the nematic phase transition occurs at the instability point: $r_{b,q}=0$ for some $q$. 
We can see that the second term in Eq.\ \eqref{rbqsmalltz} is positive and monotonically decreasing function in $q\in[0,\pi]$. As a result, the inverse susceptibility $r_{b,q}$ always has the minimum at $q=\pi$ corresponding the leading instability for the system. This indicates that the purely electronic Hamiltonian considered in this section favors the bulk nematic order with alternating sign across different layers. This $XY$ alternating order
has broken symmetry with odd and even layers having enhanced density in $X$- and $Y$-pockets correspondingly. Such alternating broken symmetry pattern is consistent to the previous finding in Ref.\ \onlinecite{PhysRevLett.102.116404}. A similar behavior also occurs in the more general model with $XY$-hybridization that we discuss in the next section.
 
 \subsubsection{Interlayer-hopping with hybridization}

The model which includes hoppings beyond the nearest neighbors is somewhat more complicated, since such hoppings make the even and odd layers not equivalent in $122$ crystal structure\cite{PhysRevB.86.144519} (see Fig.\ \ref{fig:FeAs}b). This doubles the unit cell in $z$ direction and reduces the size of the Brillouin zone by half. The momentum space representation for the next-nearest hopping \eqref{Htun'} and XY-hybridization \eqref{Hybpp} terms are
\begin{equation*}
\begin{split}
\mathcal{H}'_{\mathrm{tun}}=&-2\sum_\alpha\sum_{\mathbf{k}k_z}s_\mathbf{k}\cos k_zc^\dagger_{\alpha,k_z\mathbf{k}s}c_{\alpha,k_z\mathbf{k}s},
\\
\mathcal{H}_{\mathrm{hyb}}=&-2i\sum_\mathbf{k}{\sum_{k_z }}'\lambda_\mathbf{k}\sin k_z(c^\dagger_{X,k_z\mathbf{k}s}c_{Y,k_z+\pi,\mathbf{k}s}\\&+c^\dagger_{Y,k_z\mathbf{k}s}c_{X,k_z+\pi,\mathbf{k}s}+h.c.),
\end{split}
\end{equation*}
where $\sum_{k_z}'$ is the summation in the reduced space with $k_z\in[-\pi,0]$.  The mean-field Hamiltonian for the bulk becomes
\begin{equation}
\mathcal{H}'_{b}\!=\!{\sum_{k_zk'_z}}'\sum_{\mathbf{k}}\Psi^\dagger_{k_z}\!(\hat{H}'_{b}\!+\!\hat{V}'_b)\Psi_{k'_z}
\!+\!\frac{S}{2u}\frac{1}{N}\sum_q|\tilde{\Delta}_q|^2,
\label{Hbulk}
\end{equation}
where $\Psi_{k_z}^\dagger=(c^\dagger_{X k_z \mathbf{k}s}, c^\dagger_{X, k_z+\pi, \mathbf{k}s}, c^\dagger_{Y k_z \mathbf{k}s}, c^\dagger_{Y,k_z+\pi, \mathbf{k}s})$, 
 and the $4\times4$ matrices are 
\begin{equation*}
\begin{split}
\hat{H}'_{b}=\begin{pmatrix}
\varepsilon^X_{\mathbf{k}}-2t_\mathbf{k}\sigma^z\cos k_z & 2\lambda_\mathbf{k}\sigma^y\sin k_z\\
2\lambda_\mathbf{k}\sigma^y\sin k_z& \varepsilon^Y_{\mathbf{k}}-2t_\mathbf{k}\sigma^z\cos k_z
\end{pmatrix}\delta_{k_z,k_z'},
\end{split}
\end{equation*}
where $t_\mathbf{k}=t_z(1+2\cos k_x+2\cos k_y)$, $\sigma^{x,y,z}$ are the 
Pauli matrices
in the $(k_z,k_z+\pi)$ space, and
\begin{equation}\label{Vb}
\hat{V}'_b=\mathscr{V}'_{k_zk'_z}=
\frac{1}{N}\begin{pmatrix}
V_{k_z,k'_z}&0\\
0&-V_{k_z,k'_z}
\end{pmatrix}
\end{equation}
with the block matrix
\begin{equation*}
V_{k_zk_z'}=
\begin{pmatrix}
\tilde\Delta_{k_z-k_z'}&\tilde\Delta_{k_z-k_z'-\pi}\\\tilde\Delta_{k_z-k'_z+\pi}&\tilde\Delta_{k_z-k'_z}
\end{pmatrix}.
\end{equation*}
Diagonalizing $\hat{H}'_b$ yields the two energy-dispersion branches in the three-dimensional space
\begin{equation}\label{eqn:epm}
\begin{split}
\varepsilon^\pm_{\mathbf{k},k_z}=\varepsilon_\mathbf{k}\pm 2\sqrt{(\tfrac{\delta_\mathbf{k}}{2}-t_\mathbf{k}\cos k_z)^2+\lambda_\mathbf{k}^2\sin^2k_z},
\end{split}
\end{equation}
where $\varepsilon_\mathbf{k}=(\varepsilon^X_\mathbf{k}+\varepsilon^Y_\mathbf{k})/2$ and $\delta_\mathbf{k}=(\varepsilon^X_\mathbf{k}-\varepsilon^Y_\mathbf{k})/2$.

The calculation for the free-energy functional of $\mathcal{H}'_b$ is completely parallel to the previous section (see Appendix \ref{app:Hinfty} for detail).
Expanding the free energy up to the second order, we obtain
\begin{equation}
\begin{split}
F'_b[\tilde{\Delta}_q]&=-\frac{1}{\beta S}\ln\text{Tr}e^{-\beta(\mathcal{H}'_b-\mu\mathcal{N})}\\
&=F'_b[0]+\frac{1}{2}\int_qr'_{b,q}|\tilde{\Delta}_q|^2,
\end{split}
\end{equation}
where the inverse susceptibility $r'_{b,q}$ in this general case is
\vspace{0.15in}
\begin{widetext}
\begin{equation}\label{rbqGen}
\begin{split}
r'_{b,q}\!=\!\frac{1}{u}\!-\!\sum_{\gamma=\pm1}\!\int_{\mathbf{k}k_z}\!&
	\left[
	\frac{(\delta_\mathbf{k}\!-2t_\mathbf{k}\cos k_z\!+\gamma\eta'_{\mathbf{k}k_z})(\delta_\mathbf{k}\!-2t_\mathbf{k}\cos(k_z+q)+\gamma\eta'_{\mathbf{k},k_z})-4\lambda^2_{\mathbf{k}}\sin k_z\sin(k_z\!+q)}{4\gamma\eta'_{\mathbf{k}k_z}[\delta_\mathbf{k}t_\mathbf{k}\!-(t^2_\mathbf{k}-\lambda^2_\mathbf{k})(\cos (k_z+q)+\cos k_z)](\cos (k_z+q)-\cos k_z)\coth\frac{\beta z^\gamma_{k_z}}{2}}\right.
		\\
		&\left.-\left(\eta'_{\mathbf{k}k_z}\to \eta'_{\mathbf{k},k_z+q}\text{ and }z^{\gamma}_{k_z}\to z^{\gamma}_{k_z+q}\right)\right]
\end{split}
\end{equation}
\end{widetext}
with $z^\pm_{k_z}=\varepsilon^{\pm}_\mathbf{k}-\mu$, and $\eta'_{\mathbf{k}k_z}=2[(\frac{1}{2}\delta_\mathbf{k}+t_\mathbf{k}\cos k_z)^2+\lambda_\mathbf{k}^2\sin^2k_z]^{1/2}$. Some special-case results for $r'_{b,q}$ (circular FS with $\delta_\mathbf{k}=0$, and $q=0,\pi$) can be found in Appendix \ref{app:Hinfty}.

\begin{figure}[htbp] 
   \centering
   \includegraphics[scale=0.6]{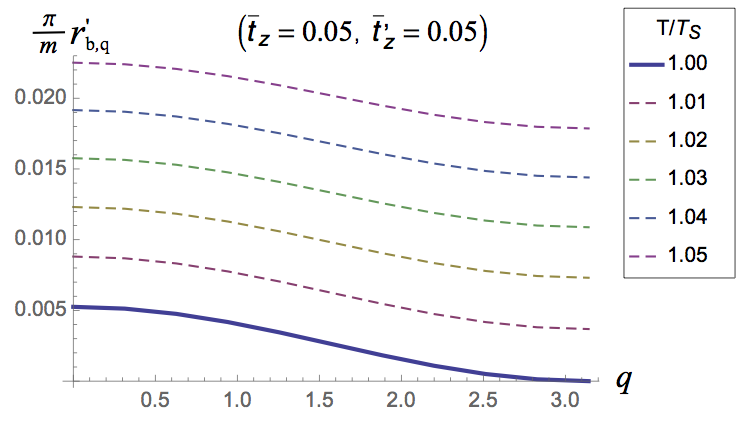} 
   \caption{The plots show the bulk inverse susceptibility $r'_{b,q}$ for different temperatures. The temperature of the thick line is the nematic transition temperature ($T_S$). The modes with $q=\pi$ has the lowest value indicating that the system favors oscillating order parameter, $\Delta_\ell=(-1)^\ell \Delta$. 
   The parameters in this plot are $m_x=m_y$, $\bar{u}=0.35, \bar{t}_z=\bar{t}'_z=0.05$.}
   \label{fig:rqvsq}
\end{figure}

To find the transition temperature, we evaluated $r'_{b,q}$ numerically. In the calculation, we let $\mu=\varepsilon_0$ be the Fermi energy and introduce the reduced parameters as follows: $\bar{\beta}=\beta \varepsilon_0$, $\bar{t}_z=t_z/\varepsilon_0$, and $\bar{u}=um/(2\pi)$.  Furthermore, we note that the inverse susceptibility in Eq.\ \eqref{rbqGen} 
is  just a linear function of the inverse coupling constant ($1/\bar{u}$). Although the transition temperature depends on the coupling constants explicitly, changing $\bar{u}$ does not change the qualitative behavior. Therefore, we use only one representative value $\bar{u}=0.35$ throughout the paper. 
Figure \ref{fig:rqvsq} shows the $q$ dependence of $r'_{b,q}$ for different temperatures.
We found that, the $q=\pi$ mode again has the smallest inverse susceptibility near the transition temperature meaning that the $XY$ alternating order also persists in this general model. However, this does not correspond to experiment, in iron pnictides the nematic order is uniform in $z$ direction. This may mean that the $q=0$ state is stabilized by external factors, such as elastic energy due to the lattice distortions induced by the nematic order. We address such stabilization below, in Section  \ref{distortion}. 

\subsection{Finite-size system and surface nematic order}

For the finite-size system, following the outline used for the bulk system in Sec. \ref{Bulk}, we begin the discussion with only nearest-neighbor hopping. This model can be solved analytically allowing us to gain some insight about the system properties. Many of these basic properties can also be found in the more realistic interlayer-hopping model. The model with $XY$ pocket hybridization has to be solved numerically, and we also discuss the method in this section.

\subsubsection{Nearest-neighbor hopping}\label{NN}
Considering only the nearest-neighbor hopping, the free energy is
\begin{equation}\label{FMatsum}
\begin{split}
F_N[&\Delta_\ell]=-\frac{1}{\beta S}\ln\text{Tr}e^{-\beta (\mathcal{H}_N-\mu\mathcal{N})}\\
&=\sum_\ell\frac{\Delta^2_\ell}{2u}\!-\!\frac{2}{\beta}\frac{1}{S}\!\sum_\alpha\!\text{tr}\ln[(\mathcal{G}^{\alpha}_{0, k})^{-1}\!+\!\mathcal{V}^\alpha].
\end{split}
\end{equation}
Here we have introduced the notation $k=(\omega_n,\mathbf{k})$, the matrix $(\mathcal{G}^{\alpha}_{0,k})^{-1}$ is defined as
\begin{equation*}
(\mathcal{G}^{\alpha}_{0,k})^{-1}=
\begin{pmatrix}
(G^\alpha_{0, k})^{-1} &-t_z&\cdots&0\\
-t_z&(G^\alpha_{0, k})^{-1} &\ddots&\vdots\\
\vdots &\ddots &\ddots &-t_z\\
0&\cdots& -t_z &(G^\alpha_{0, k})^{-1}
\end{pmatrix},
\end{equation*}
where $(G^\alpha_{0, k})^{-1}=i\omega_n-\xi^\alpha_{\mathbf{k}}$ is the in-plane one-particle Green's function, and $[\mathcal{V}^\alpha]_{\ell\ell'}=V^\alpha_{\ell}\delta_{\ell\ell'}$ is the diagonal matrix of order parameters.

To find the critical point, we have to expand the free energy, Eq.\ \eqref{FMatsum}, with respect to $\Delta_\ell$ up to the second order,
\[F_N[\Delta_\ell]\simeq F_N[0]+\frac{1}{2}\sum_{\ell\ell'}r_{\ell\ell'}\Delta_\ell\Delta_{\ell'}.\]
The inverse susceptibility is
\begin{equation}\label{rll0}
r_{\ell\ell'}=\frac{\delta_{\ell\ell'}}{u}+\frac{2}{\beta}\sum_{\omega_n,\alpha}\int_\mathbf{k}[\mathcal{G}^{\alpha}_{0}]_{\ell\ell'}[\mathcal{G}^{\alpha}_{0}]_{\ell'\ell},
\end{equation}
where $[\mathcal{G}^{\alpha}_{0}]_{\ell\ell'}$ is the tight-binding Green's function in the layer-index basis (see Appendix \ref{app:Ginverse}),
\begin{equation}\label{eqn:Galpha-1}
[\mathcal{G}^{\alpha}_{0}]_{\ell\ell'}=\sum_{p}\frac{2/(N+1)\sin \ell\vartheta_{p}\sin \ell'\vartheta_p}{i\omega_n-\varepsilon^\alpha_{\mathbf{k}}+\mu+2t_z\cos\vartheta_p}
\end{equation}
with $\vartheta_p=\frac{p\pi}{N+1}$ and $p=1\dots N$.

Carrying out the Matsubara frequency summation in Eq.\ \eqref{rll0} explicitly, we obtain
\begin{equation}\label{rll}
\begin{split}
r_{\ell\ell'}=\frac{\delta_{\ell\ell'}}{u}-\sum_{p,\alpha}\int_\mathbf{k}\Big[\frac{\beta}{2}(S^p_{\ell\ell'})^2\mathrm{sech}^2\Big(\frac{\beta z^\alpha_p}{2}\Big)\\-\frac{1}{2}\sum_{p'\neq p}S^p_{\ell\ell'}S^{p'}_{\ell'\ell}\frac{\tanh\frac{\beta z^\alpha_p}{2}-\tanh\frac{\beta z^\alpha_{p'}}{2}}{2t_z(\cos\vartheta_p-\cos\vartheta_{p'})}\Big].
\end{split}
\end{equation}
where $S^p_{\ell\ell'}=2\sin \ell\vartheta_p\sin \ell'\vartheta_p/(N+1)$, and $z^\alpha_p=\varepsilon^\alpha_{\mathbf{k}}-\mu-2t_z\cos\vartheta_p$. The quadratic matrix $r_{\ell\ell'}$ in Eq. \eqref{rll} contains the essential information for analyzing the nematic phase transition at the critical point. 

Expanding Eq.\ \eqref{rll} with respect to $t_z$ and performing integration with respect to $\mathbf{k}$ (see Appendix \ref{smalltz}), we obtain
\begin{align}
		r_{\ell\ell'}\!&\simeq
		r_1\delta_{\ell\ell'}\nonumber\\
		+&r_t\Big[\delta_{\ell\ell'}\!+\!\frac{\delta_{\ell,\ell'+1}\!+\delta_{\ell,\ell'-1}}{4}\!-\!\frac{\delta_{\ell,1}\!+\delta_{\ell,N}}{2}\delta_{\ell\ell'}\Big].
		\label{eq:smalltz} 
\end{align}
The first line is just the single-layer term derived before, Eq.\ \eqref{r1result},  and the second line is the correction from the interlayer tunneling. 
The first two terms in the second line describe bulk interactions and correspond to previous derivation in the momentum space, Eqs.\ \eqref{rbqsmalltz} and \eqref{rt}.
As expected, the lowest-order expansion with respect to $t_z$ gives only interaction between $\Delta_\ell$ in neighboring layers.
The coefficients $r_{\ell\ell'}$ with $|\ell-\ell'|>1$ correspond to the higher-order corrections in $t_z$. This implies that the nematic order parameters do not have long-range interactions in the out-of-plane direction. The third term in the interlayer contribution represents  the surface correction. Its negative sign implies that the surface favors the nematic order. Thus, in this model, surface transition should be expected. These properties, derived analytically from the nearest-neighbor model, also preserve in the more realistic model that includes the hoppings beyond the nearest neighbor, which we consider in the next section. 

\subsubsection{Inter-layer hopping with hybridization}

For completeness, we also consider hopping processes beyond the nearest neighbor. 
Taking into account tunneling terms described by Eqs.\ \eqref{Htun'} and \eqref{Hybpp},
the free energy becomes
\begin{equation*}
\begin{split}
F'_N[\Delta_\ell]=-\frac{1}{\beta S}\ln\text{Tr}e^{-\beta (\mathcal{H}_N+\mathcal{H}'_{\mathrm{tun}}+\mathcal{H}_{\mathrm{hyb}}-\mu\mathcal{N})}.
\end{split}
\end{equation*}
As the $N$-layers Hamiltonian is just a one-body field operator, the free energy can be immediately written down as
\begin{equation}\label{FNE}
F'_N=\sum_\ell\frac{\Delta^2_\ell}{2u}-\frac{2}{\beta}\sum_{p}\int_\mathbf{k}\ln \Big[1+e^{-\beta f_{p,\mathbf{k}}[\Delta_\ell]}\Big],
\end{equation}
where $f_{p,\mathbf{k}}$ is the quasiparticle energy, which are the $p$-th eigenvalue ($p=1\dots 2N$) of the following Hamiltonian matrix
\begin{equation}\label{Hhat}
\hat{H}=
\begin{pmatrix}
\mathbf{E}^X_{\mathbf{k}} & \mathbf{P}_\mathbf{k}\\
\mathbf{P}_\mathbf{k} &\mathbf{E}^Y_{\mathbf{k}}
\end{pmatrix}+
\begin{pmatrix}
\mathcal{V}^X & 0\\
0 &\mathcal{V}^Y
\end{pmatrix}.
\end{equation}
The block matrices are
\begin{subequations}
\begin{equation}\label{Ek}
\mathbf{E}^\alpha_\mathbf{k}=
	\begin{pmatrix}
	\varepsilon^\alpha_{\mathbf{k}}-\mu & -t_\mathbf{k}  & \cdots & 0\\ 
	-t_\mathbf{k} & \varepsilon^\alpha_{\mathbf{k}}-\mu&\ddots&\vdots\\
	\vdots & \ddots  &\ddots&-t_\mathbf{k} \\
	0 & \cdots &-t_\mathbf{k}& \varepsilon^\alpha_{\mathbf{k}}-\mu
	\end{pmatrix},
\end{equation}
with $t_\mathbf{k}=t_z(1+2\cos k_x+2\cos k_y)$,
\begin{equation}\label{Pk}
\mathbf{P}_\mathbf{k}\!=\!
	\lambda_{\mathbf{k}}\!
	\begin{pmatrix}
	0 & (\!-\!1)^1  & \cdots  & 0\\ 
	(\!-\!1)^1 & 0&\ddots&\vdots\\
	\vdots & \ddots  &\ddots&(\!-\!1)^{N\!-\!1} \\
	0 & \cdots &(\!-\!1)^{N\!-\!1}& 0
	\end{pmatrix},
\end{equation}
and $[\mathcal{V}^\alpha]=V^\alpha_\ell\delta_{\ell\ell'}$.
\end{subequations}

%
Due to the off-diagonal block matrix $ \mathbf{P}_\mathbf{k}$ in $\hat{H}$, there is no simple analytical expression for the free energy in this case. To evaluate the free energy, we use Eq.\ \eqref{FNE} and solve the eigenvalues problem numerically. 
The eigenvalues ($f_{p,\mathbf{k}}$ ) of $\hat{H}$ can be calculated by treating the second term in Eq.\ \eqref{Hhat} as a perturbation, since $\Delta_\ell$ are small at the critical point. To expand the eigenvalues at $\Delta_\ell\simeq0$,
we first solve for the eigenvalues and eigenvectors of the first term in Eq.\ \eqref{Hhat}, which we notate as 
\[
f^{(0)}_{p,\mathbf{k}} ,\text{ and } \mathbf{x}^T_{p,\mathbf{k}}=(x^1_{p,\mathbf{k}},\dots,x^{2N}_{p,\mathbf{k}}) 
\]
respectively. The eigenvectors are normalized as $\mathbf{x}_p^T\mathbf{x}_p=1$.

We apply perturbation expansion to $f_{p,\mathbf{k}}[\Delta_\ell]$. If $N$ is even or odd with $\lambda_\mathbf{k}\neq0$ and $m_x\neq m_y$, all $f^{(0)}_{p,\mathbf{k}}$ are distinct and non-degenerate. Therefore, the approximate eigenvalues are
\begin{equation}\label{2ndpert}
\begin{split}
f_{p,\mathbf{k}}&\simeq f^{(0)}_{p,\mathbf{k}}+\mathbf{x}^T_p\hat{V}\mathbf{x}_p+\sum_{p'\neq p}\frac{(\mathbf{x}^T_p\hat{V}\mathbf{x}_{p'})^2}{f^{(0)}_{p,\mathbf{k}}-f^{(0)}_{p',\mathbf{k}}},\\
&=f^{(0)}_{p,\mathbf{k}}+\sum_\ell a^p_\ell\Delta_\ell+\sum_{\ell\ell'}b^p_{\ell\ell'}\Delta_\ell\Delta_{\ell'},
\end{split}
\end{equation}
where $\hat{V}$ is the second term in Eq.\ \eqref{Hhat}, and  
\begin{equation}
a^p_\ell=v^\ell_{pp}, \quad
b^p_{\ell\ell'}=\sum_{p'\neq p}\frac{v^\ell_{pp'}v^{\ell'}_{pp'}}{f^{(0)}_{p,\mathbf{k}}-f^{(0)}_{p',\mathbf{k}}}\label{ab}
\end{equation}
with $v^\ell_{pp'}=x^\ell_{p,\mathbf{k}}x^\ell_{p',\mathbf{k}}-x^{N+\ell}_{p,\mathbf{k}}x^{N+\ell}_{p',\mathbf{k}}$.
Therefore, using Eq.\ \eqref{2ndpert} near $\Delta_\ell\simeq0$ and
expanding the free energy, we obtain the inverse susceptibilities with hybridization,
\begin{equation}\label{rll'}
\begin{split}
r'_{\ell\ell'}=\frac{\delta_{\ell\ell'}}{u}-\sum_{p}\int_\mathbf{k}\Big[\frac{\beta}{2}a^p_\ell a^p_{\ell'}\mathrm{sech}^2\Big(\frac{\beta z_p}{2}\Big)\\+2b^p_{\ell\ell'}\tanh\Big(\frac{\beta z_p}{2}\Big)\Big],
\end{split}
\end{equation}
where $z_p=f^{(0)}_{p,\mathbf{k}}-\mu$. To facilitate the integration over the in-plane momentum $\mathbf{k}$, we approximate the energy dispersion near the FS as
\begin{equation}
\varepsilon^{X,Y}_\mathbf{k}\simeq\frac{\mathbf{k}^2}{2m}\pm\delta_2\cos2\theta,
\end{equation} 
where $m=2m_xm_y/(m_x+m_y)$ and $\delta_2=\varepsilon_0(1-m_x/m_y)/2$. The upper `$+$' (lower `$-$') sign is for the $X$ ($Y$) pocket electrons. Furthermore, the momentum integration can be done by using $\int_\mathbf{k}=\frac{m}{2\pi}\int^{\mu+\epsilon_c}_0d\varepsilon\int^{2\pi}_{0}\frac{d\theta}{2\pi}$, where $\epsilon_c$ is some cutoff energy of the model with the scale of bandwidth energy.

The case when $N$ is odd and $m_x=m_y$ requires special consideration in numerical calculations, see Appendix \ref{degen}. In this case the eigenspace of the first term in Eq.\ \eqref{Hhat} breaks into $N$ 2-fold degenerate subspaces meaning that the formula in Eq.\ \eqref{ab} has to be modified. 

\subsubsection{Calculation of transition temperature}\label{result}

We will use the same notations for the reduced parameters as in Sec. \ref{Bulk}, i.\ e., $\bar{\beta}=\beta \varepsilon_0$, $\bar{t}_z=t_z/\varepsilon_0$,  $\bar{u}=um/(2\pi)$, and, also, $\bar{\delta}_2=\delta_2/\varepsilon_0$.
\begin{figure*}[htbp]
   \centering
   \includegraphics[scale=0.63]{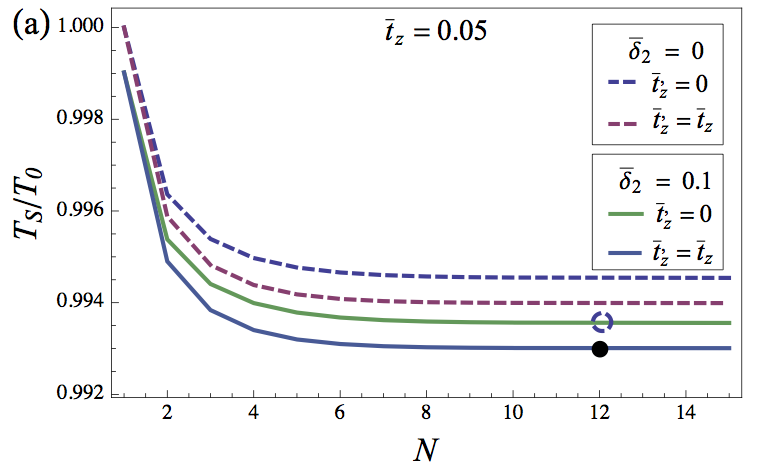}  
   \includegraphics[scale=0.63]{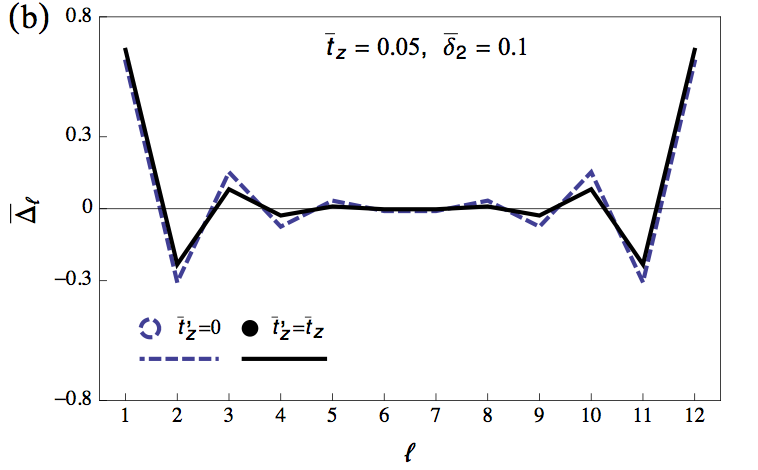}
   \caption{(a) The plot of the nematic transition temperature $T_S$ versus the number of layers in the system. $T_0$ is the transition temperature of the single layer at $\mu=\varepsilon_0$ with $\bar{u}=0.35$. This plot indicates that the interlayer hoppings  (both $t$ and $t'_z$) always lower the system transition temperature. Note that due to the $\mathbf{k}$-dependent in the hopping terms ($t_\mathbf{k}$ and $\lambda_\mathbf{k}$), the FS ellipticity ($\delta_2$) also influences the $T_S$. (b) The spatial configuration of the lowest eigenmode near $T_S$ for the system with $N=12$. The dashed (solid) line correspond to the point in (a) that are marked by the open circle (filled circle). The order parameter has the maximum at the surface and decays in the bulk. We have defined $\bar{\Delta}_\ell=\Delta_\ell/\sqrt{\sum_{\ell}\Delta^2_\ell}$.
   }
   \label{fig:TCvsN}
\end{figure*}
To obtain the transition temperature,  $T_S$, we search for the $\bar{\beta}=\varepsilon_0/T_S$ such that the lowest eigenvalue of $r'_{\ell\ell'}$, Eq.\ \eqref{rll'}, approaches zero meaning that at $T_S$ the equation $\sum_{\ell'=1}^Nr'_{\ell,\ell'}\Delta_{\ell'}=0$ has a nontrivial solution $\Delta_{\ell'}\neq 0$. We examine evolution of the transition temperature with increasing number of layers $N$.

For the nearest-neighbor model at $t'_{z}\!=\!0$ and small $t_{z}$, when the inverse susceptibility is given by Eq.\ \eqref{eq:smalltz}, the problem has simple analytical solution for $N\!\gg \! 1$. Near the surface $\ell\! =\!0$ we obtain the following system 
\begin{equation*}
\left( r_{1}+r_{t}\right) \Delta _{\ell }+\frac{r_{t}}{4}\left( \Delta
_{\ell -1}+\Delta _{\ell +1}\right) -\frac{r_{t}}{2}\delta _{\ell ,1}\Delta
_{1}=0,
\end{equation*}
for $\ell \geq 1$ and $\Delta _{0}=0$. Looking for solution in the form $\Delta _{\ell }\propto (-1)^{\ell }\exp (-\varkappa \ell )$, we obtain
\begin{subequations}
\begin{align}
r_{1}+r_{t}-\frac{r_{t}}{2}\cosh \varkappa  &=0\text{, for }\ell >1,
\label{TsEq1} \\
r_{1}+\frac{r_{t}}{2}-\frac{r_{t}}{4}\exp (-\varkappa ) &=0\text{, for }\ell =1.  \label{TSeq2}
\end{align}
\end{subequations}
These two equations yield 
\begin{equation}
\exp \varkappa =2  \label{DecayLengthNN}
\end{equation}
and equation
\begin{equation}
r_{1}+\frac{3r_{t}}{8}=0,  \label{eqTS-NN}
\end{equation}
which determines the surface instability temperature for the nearest-neighbor model.
This result has to be compared with the bulk-transition equation, \[r_{1}+\frac{r_{t}}{2}=0,\] which can be obtained by setting $\varkappa =0$ in Eq.\ \eqref{TsEq1}.  

In general case with arbitrary $t_z$ and $t'_z$ we solved equations for $T_S$ numerically.
Fig. \ref{fig:TCvsN}a shows representative dependences $T_S(N)$ obtained for $\bar{t}_z=0.05$ and different $\bar{t}_{z}^\prime$ and $\bar{\delta}_2$.
The transition temperature decreases as more layers are added to the system. For very large $N$, $T_S$ eventually approaches a finite limiting value. Furthermore, $T_S$ always decreases as the hopping energies $t_z$ and $t_z'$ increase. This implies that the hopping between layers suppresses the  nematic order. 

\begin{figure}[htbp] 
   \centering
   \includegraphics[scale=0.65]{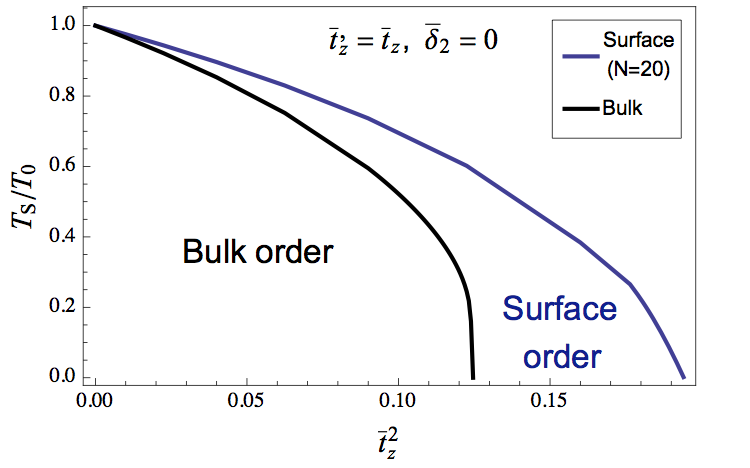}
   \caption{This plot shows the $t_z$ dependence of the bulk and surface transition temperatures.   Both transitions are suppressed by the interlayer tunneling and nematic order can be completely destroyed if $t_z$ is too large. The surface-nematic range rapidly increases with increasing $t_z$. Within some range of $t_z$ only the surface nematic state exists.
   }
   \label{fig:TStz}
\end{figure}

Near the transition points, the sign change in the eigenvalue of $r'_{\ell\ell'}$ also indicates the divergence in nematic susceptibility for the corresponding eigenmode, which signals an instability of this eigenmode. Examining the lowest eigenmode with zero eigenvalue at $T_S$ allows us to deduce the most energetically favorable nematic spatial configuration. 
Figure 
 \ref{fig:TCvsN}b shows coordinate dependence of the unstable eigenmode for different parameters. We see that this eigenmode  decays away from the surface so that for sufficiently large $N$, nematic order becomes vanishingly small at the center. This result implies that instability for finite-size systems at large $N$ corresponds to formation of surface nematic and limiting values of $T_S$ at large $N$ in Fig. \ref{fig:TCvsN}a correspond to surface instability. In Fig.\ \ref{fig:TStz} we compare $t_z$-dependences of the bulk and surface transition temperatures. The split between these transitions increases with $t_z$. 
In addition, if $t_z$ is too strong, the nematic phase transition disappears. In this large $t_z$ case, the lowest eigenvalue of $r'_{\ell\ell'}$ never becomes negative and the symmetry unbroken phase, $\Delta_\ell=0$, always remains the true global minimum of $F_N[\Delta_\ell]$.
Within some range of $t_z$ only the surface nematic exists without bulk transition.

\begin{figure}[htbp] 
   \centering
   \includegraphics[scale=0.61]{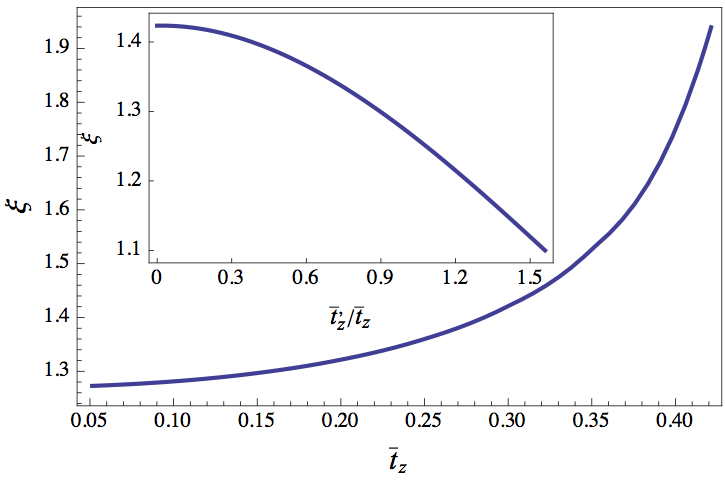}
   \caption{The decay length of surface nematic obtained using system with $N=12$ for $\bar{t}'_z=0.05$. Note that, the  nematic order vanishes for $\bar{t}_z>0.44$. The inset shows $\bar{t}'_z$ dependence of $\xi$ for fixed $\bar{t}_z=0.05$.
   }
   \label{fig:decay}
\end{figure}

To see how spatial configuration of the nematic order parameter depends on different hopping effects, we fit the nematic order parameter from the surface ($1\leq\ell<N/2$) to the exponential function,
\begin{equation}\label{expfit}
|\bar{\Delta}_\ell|=\Delta_0 e^{-\ell/\xi}
\end{equation}
where $\xi$ is the decay length in units of the interlayer spacing, and $\Delta_0$ is some constant.
Figure \ref{fig:decay} illustrates dependence of the decay length at $T_S$ on the interlayer hopping parameters. We can see that
the decay length is typically very small, only 1-2 interlayer spacings.
It increases with increasing $\bar{t}_z$, but decreases with increasing $\bar{t}'_z$ (see the inset). Note that the value of $\xi$ at $\bar{t}'_z=0$ reproduces the value $1/\varkappa=1/\ln(2)\approx 1.44$ analytically derived above, Eq.\ \eqref{DecayLengthNN} .

\section{The effects from lattice distortion}\label{distortion}

We found that the interlayer interactions mediated by the electronic tunneling favor $XY$-alternating nematic order. 
However, such order is not favorable for the lattice elastic energy. Indeed, if one layer is stressed in the $X$-direction and its neighboring layer is stressed in the $Y$-direction, this distortion increases the interlayer ion-ion distances in lattice. Thus, this costs higher elastic energy than stressing all layers in the same direction. Therefore, in the electron-lattice coupled system, $XY$-alternating order may not yield the lowest free energy.

For quantitative treatment of this problem, we consider the free energy with the following simple extension
\begin{equation}\label{FNel}
\mathcal{F}_N[\Delta_\ell,u_\ell]=F'_N[\Delta_\ell]+F_{el}[\Delta_\ell,u_\ell],
\end{equation}
where the elastic part is modeled by
\begin{equation}\label{eqn:Fel}
F_{el}[\Delta_\ell,u_\ell]=-g\sum_\ell u_\ell\Delta_\ell +\frac{1}{2}\sum_{\ell\ell'}C_{\ell,\ell'}u_\ell u_{\ell'}.
\end{equation}
Here $u_\ell=\frac{a_\ell-b_\ell}{a_\ell+b_\ell}$ is the $\ell$-th layer lattice
distortion, and $a_\ell$ and $b_\ell$ are the in-plane lattice constant in $x$- and $y$-
direction respectively. $C_{\ell,\ell'}$ is the shear modulus constants, and $g$ is the
coupling between the nematic order parameter and lattice distortion. We will neglect
temperature dependences of these parameters.  For simplicity, we only consider the elastic
matrix up to nearest neighbor and use notations $C_{\ell,\ell}=C_s$ and  $C_{\ell,\ell\pm
	1}=-C'_s$. We assume  $C_{\ell,\ell\pm 1}$ to be negative so that it is favorable for
layers in the lattice to be stressed in the same direction. The bulk shear modulus is
given by $C_{66}=(C_s-2C'_s)/c_z$ with $c_z$ being the c-axis lattice parameter meaning that
the elastic constants must satisfy the condition $C_s/C_s'>2$ in order to have a stable
lattice.

\subsection{Bulk nematic transition}

For the $N\to \infty$ bulk limit, the free energy of the elastic part in the momentum space is
\begin{equation*}
F_{el}=-g\int_q \tilde{\Delta}_{q}\tilde{u}_{-q}+\frac{1}{2}\int_q(C_s-2C_s'\cos q)\tilde{u}_q\tilde{u}_{-q}
\end{equation*}
where $\tilde{u}_q=\sum^\infty_{\ell=-\infty}e^{iq\ell}u_\ell$. 
Minimizing $F_{el}$ with respect to $u_{-q}$, we obtain
\begin{equation}\label{elasticF2}
\tilde{u}_q=\frac{g\tilde{\Delta}_q}{C_s-2C_s'\cos q}.
\end{equation}
Therefore, the optimal elastic free energy is
\begin{equation}
F_{el}=-\frac{1}{2}\int_q\frac{g^2|\tilde{\Delta}_q|^2}{C_s-2C'_s\cos q}.
\end{equation}
This new negative term modifies the inverse susceptibility as
\begin{equation}
\bar{r}_{b,q}=r'_{b,q}-\frac{g^2}{2(C_s-2C'_s\cos q)}.
\end{equation} 
The elastic correction always reduces the transition temperature. Moreover, one can check that the elastic correction reduces the value of $\bar{r}_{b,0}$ more than $\bar{r}_{b,\pi}$. This implies that with the elastic correction $\bar{r}_{b,\pi}$ may not be the minimum inverse susceptibility any more. For sufficiently strong coupling $g>g_{c1}$, $\bar{r}_{b,0}$ drops below $\bar{r}_{b,\pi}$ at the transition temperature and the system starts to favor a uniform order. The condition for the critical coupling strength is determined by
\begin{equation}\label{con:gc1}
\bar{r}_{b,0}|_{T=T_S}=\bar{r}_{b,\pi}|_{T=T_S}=0,
\end{equation}
giving
\begin{equation}\label{cgc1}
\frac{2g_{c1}^2/C'_s}{(C_s/C'_s)^2-4}+r'_{b,0}|_{T=T_S}-r'_{b,\pi}|_{T=T_S}=0.
\end{equation}

In particular, for the model with only nearest-neighbor hopping in the small-$t_z$ limit, the inverse susceptibility has simple analytical form, Eq.\ (\ref{rbqsmalltz}). 
In this case Eq.\ \eqref{cgc1} simply becomes
\begin{equation}
\frac{2g_{c1}^2/C'_s}{(C_s/C'_s)^2-4}+r_t|_{T=T_S}=0,
\end{equation}
where $r_t$ is defined in Eq.\ \eqref{rt}. 

\begin{figure}[htbp]
   \centering
   \includegraphics[scale=0.46]{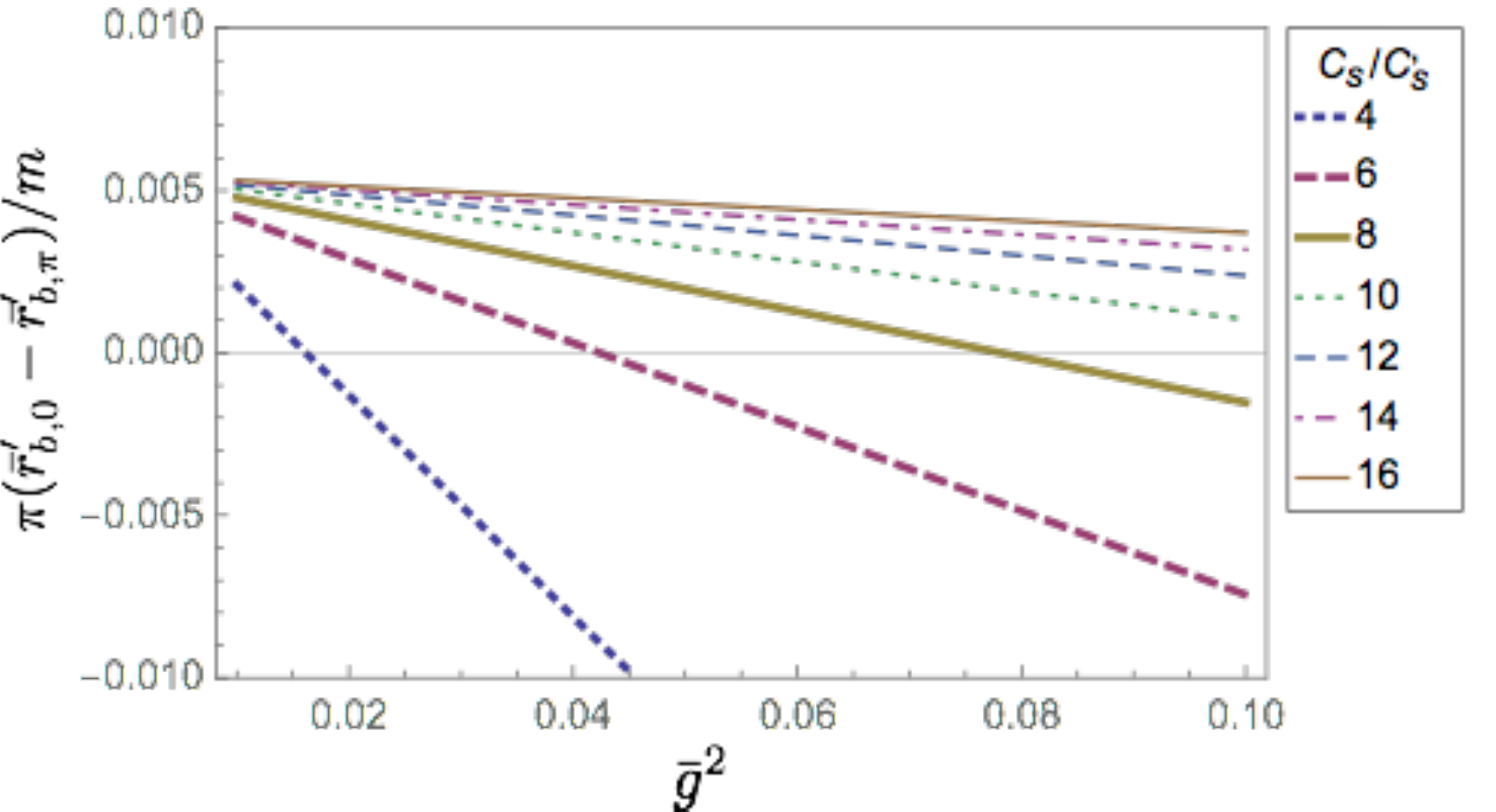}
      \caption{The $\bar{g}^2$ dependence of the inverse susceptibility $r^b_q$ at the  temperature with $r^b_\pi=0$. At some $\bar{g}=\bar{g}_{c1}$, the difference 
      	$r^b_{q=0}-r^b_{q=\pi}$ becomes negative at the transition temperature. This indicates that  the uniform bulk nematic order becomes more favorable than the $XY$-alternating order. The plots are made using $\bar{u}=0.35$, $\bar{t}_z=0.05$, and $\bar{t}'_z=0$.}
   \label{fig:Nelastic}
\end{figure}

For numerical analysis we introduce the reduced parameters $\bar{g}=g\sqrt{\pi / m C_s'}$, and $C_s/C'_s$. Figure \ref{fig:Nelastic} illustrates dependences of the difference $\bar{r}_{b,0}-\bar{r}_{b,\pi}$ at the transition point on the reduced coupling strength $\bar{g}^2$ for different ratios $C_s/C'_s$. The zero-crossing of these plots determines the critical coupling strength $g_{c1}$. We can see that it rapidly increases with $C_s/C'_s$. 
For more realistic model, which takes into account $XY$-pocket hybridization, the critical value of coupling $g_{c1}$ has to be found numerically from Eq.\ \eqref{cgc1} using full expression for the electronic inverse susceptibility $r_{b,q}'$, Eq.\ \eqref{rbqGen}.  

\subsection{Finite-size system and surface-nematic transition}
For the finite-size system, to find the nematic order ground state, we minimize the free energy with respect to $u_\ell$. 
\[
\frac{\delta \mathcal{F}}{\delta u_\ell}=\frac{\delta F_{el}}{\delta u_\ell}=0.
\]
This yields a system of linear equations
\begin{equation}\label{eqn:dF0}
-g\Delta_\ell+C_su_\ell-C'_s(u_{\ell+1}+u_{\ell-1})=0
\end{equation}
with $u_0=u_{N+1}=0$. Solving the equation by inverting the Toeplitz tridiagonal matrix,\cite{0305-4470-29-7-020} we obtain the required $u_\ell$ which minimizes $F_{el}$, 
\begin{equation}
u_\ell=\frac{g}{C'_s}\sum_{\ell'}M_{\ell\ell'}\Delta_{\ell'}
\end{equation}
where the matrix $M_{\ell\ell'}$ is
\begin{equation*}
\begin{split}
M_{\ell\ell'}=\frac{\cosh(\kappa\varphi^-_{\ell\ell'})-\cosh(\kappa\varphi^+_{\ell\ell'})}{2\sinh \kappa\sinh [(N+1)\kappa]}
\end{split}
\end{equation*}
with $\varphi^\pm_{\ell\ell'}=N+1-|\ell\pm\ell'|$ and $\cosh \kappa=C_s/(2C_s')$. Substituting these lattice distortions into $F_{el}$, we immediately obtain the optimal free energy as
\begin{equation}\label{Felmin}
F_{el}=-\sum_{\ell\ell'}\frac{g^2}{2C'_s}M_{\ell\ell'}\Delta_\ell\Delta_{\ell'}.
\end{equation}

\begin{figure*}[htbp] 
   \centering
   \includegraphics[scale=0.335]{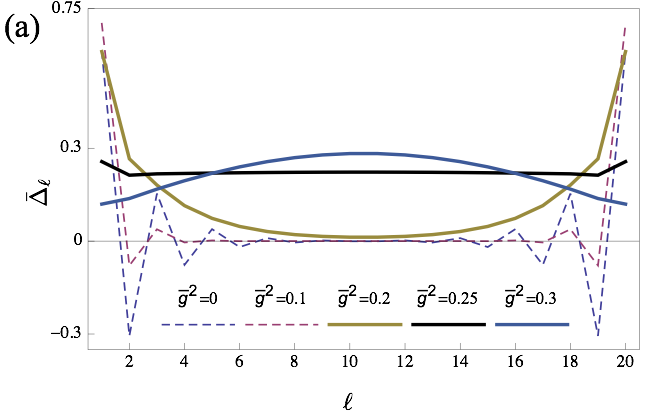} 
   \includegraphics[scale=0.385]{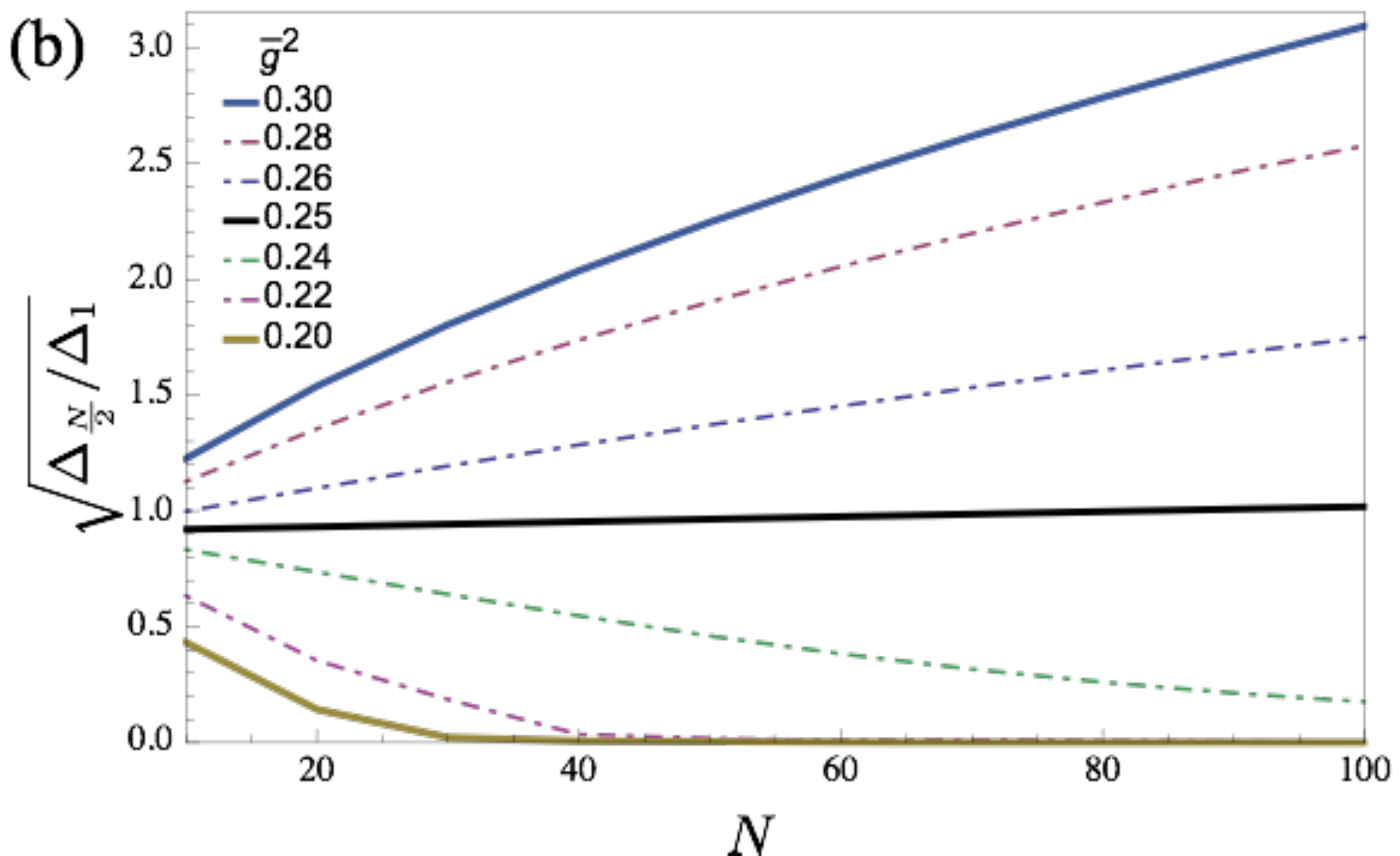} 
   \caption{(a) Spatial configuration of the nematic order with the effects of lattice distortion for $N=20$, $\bar{t}_z=0.05$, $\bar{t}'_z=0$,  $C_s/C'_s=10$,  and different coupling strengths $\bar{g}^2$. Depending on the coupling strength $\bar{g}$, the nematic order cease to oscillate across different layers at $\bar{g}^2\simeq 0.2$. For $\bar{g}^2>0.25$, the instability corresponds to bulk transition. (b) $N$ dependence of the order parameter at the center for three values of $\bar{g}$ close to 
   $\bar{g}_{c2}$. For  $\bar{g}^2\lesssim0.25$ the nematic order at the center approaches zero with increasing $N$ corresponding to the surface order. The instability is bulk for $\bar{g}^2\gtrsim0.25$. 
   }
   \label{fig:elastic}
\end{figure*}

The optimal spatial configuration for the nematic order can be determined using Eqs.\ \eqref{FNel} and \eqref{Felmin}. We remark that the coupling to lattice leads to higher transition temperature. This indicates that lattice distortion actually promotes the formation of nematic order. Evolution of spatial dependence of the order parameter with increasing coupling strength $\bar{g}$ is shown in Fig.\ \ref{fig:elastic}a for $N=20$. Other parameters are  $\bar{u}=0.35$, $\bar{t}_z=0.05$, $\bar{t}'_z=0$, and $C_s/C'_s=10$. 
We can see that coupling to the lattice distortion can change the spatial configuration of the nematic order drastically. If the electron-lattice coupling $\bar{g}$ is large enough, the order parameter no longer oscillates across different layers. On the other hand, the coupling to the lattice suppresses surface instability.
When the coupling strength $\bar{g}$ exceeds certain critical value, $\bar{g}_{c2}$, the intermediate surface nematic disappears and only bulk transition remains, see, e.~g., the plot for $\bar{g}^2=0.3$ in Fig. \ref{fig:elastic}a. Finding this critical value requires careful study of finite-size effects for very large $N$. Fig.\ \ref{fig:elastic}b shows the size dependence of the ratio of nematic order parameters at the center, $\ell=N/2$, and at the surface, $\ell=1$, for seven values of $\bar{g}$ close to $\bar{g}_{c2}$. We can see that for $\bar{g}<0.25$ this ratio decays to zero with increasing $N$ (surface order) while for $\bar{g}\geq 0.25$ it grows (bulk order).\footnote{One can demonstrate that if the order parameter is suppressed at the surface then at the transition point the ratio $\Delta_{N/2}/\Delta_{1}$ grows linearly with the system size $N$. We can expect that below the transition temperature this ratio becomes size independent at large $N$. } This means that $0.24<\bar{g}_{c2}^2<0.25$.

\begin{figure}[htbp]
   \centering
   \includegraphics[scale=0.46]{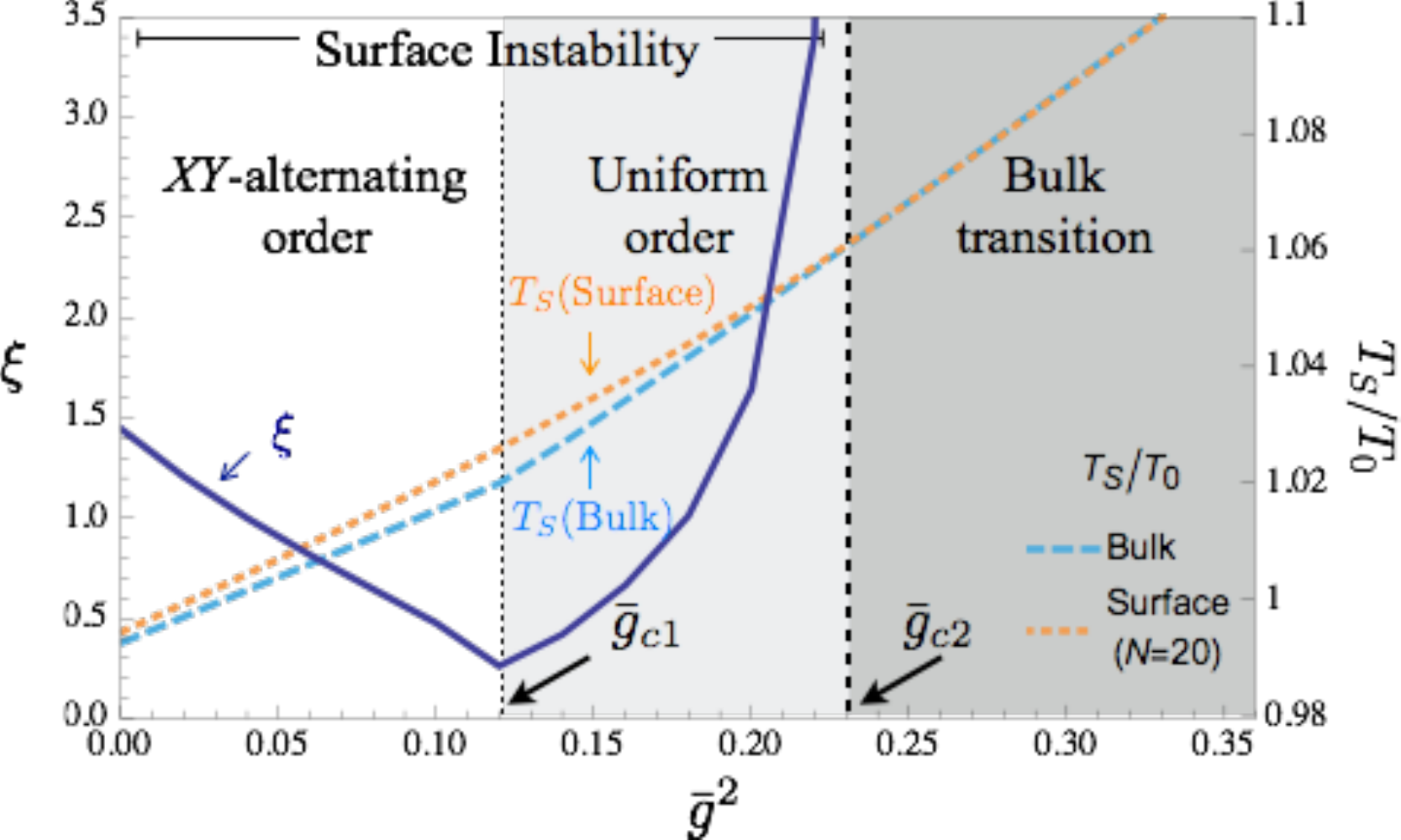} 
   \caption{
The decay length (solid line) and the transition temperature (dotted line) of the system with $N=20$, $\bar{t}_z=0.05$, $\bar{t}'_z=\bar{\delta}_2=0$, and $C_s/C'_s=10$. The dashed line is the transition temperature of the bulk system ($N=\infty$). The kink is approximately located at $\bar{g}^2_{c1}\simeq0.12$, which is the transition point from XY-alternating order to uniform order.The decay length diverges at $\bar{g}^2_{c2}\simeq0.25$ and only bulk transition remains after this point.
   }
   \label{fig:decayg2}
\end{figure}
\begin{figure}[htbp]
	\centering
	\includegraphics[scale=0.42]{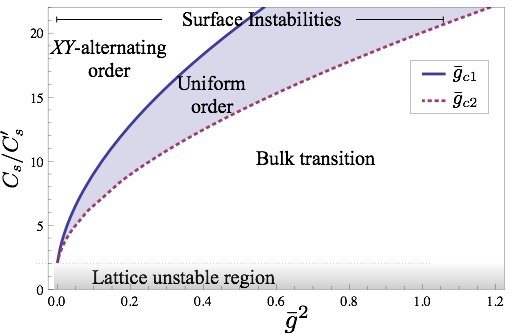}
	\caption{
		This diagram shows three regions with different transition behaviors in the parameter plane $\bar{g}^2 - C_s/C'_s$. The shaded region
		corresponds to existence of surface instability for uniform bulk ground state. 
	}
	\label{fig:SI}
\end{figure}
To further characterize the influence of lattice distortion on surface nematic, we computed its decay length by fitting the order parameters with Eq.\ \eqref{expfit}. 
Figure\ \ref{fig:decayg2} shows the dependence of the decay length, $\xi$, on the reduced coupling constant. We can see that this length is a nonmonotonic function of $\bar{g}^2$; it decreases with  $\bar{g}^2$ for $\bar{g}<\bar{g}_{c1}$ and increases with  $\bar{g}^2$ for  $\bar{g}>\bar{g}_{c1}$. A very small minimum value, $\xi\approx 0.3$ is realized at the transition point from the alternating to uniform order, $\bar{g}=\bar{g}_{c1}$, where the dependence has a kink.\footnote{Note, however, that in the vicinity of $\bar{g}_{c1}$ the decay of the order parameter is not well described by a simple exponent, Eq.\ \eqref{expfit}.} The decreasing $\xi(\bar{g})$ in the alternating-order region can be easily understood.
$\xi^2$ is proportional to the stiffness  $r''=d^2r_{b,q}/dq^2|_{q=q_0}$, where $q_0=0,\pi$ is the ground-state wave vector. 
The lattice contribution to $r_{b,q}$ has minimum at $q=0$ and maximum at $q=\pi$. Therefore, in the alternating state ($q_0=\pi$) the electronic part of $r''$ is positive and the lattice contribution is negative meaning that the increase of $\bar{g}$ reduces the stiffness $r''$ and, correspondingly,  decreases $\xi$. In the uniform state ($q_0=0$) the growth of the lattice contribution enhances stiffness at $q_0=0$ and increases $\xi$. We can see that the decay length diverges rapidly as $\bar{g}$ approaching $\bar{g}_{c2}$ from below, where the surface-nematic state disappears. 

Figure\ \ref{fig:decayg2} also shows the coupling-constant dependences of the surface and bulk transition temperatures. We can see that the bulk $T_S$ has a kink at $g=\bar{g}_{c1}$ while the surface transition is smooth at this point.  The kink in bulk $T_S$ is a natural consequence of qualitative change of the ground-state configuration from alternating to uniform order. At $\bar{g}=\bar{g}_{c2}$ the surface transition smoothly merges with the bulk one. For $\bar{g}>\bar{g}_{c2}$ the transition temperature evaluated for finite-size system with $N=20$ is indistinguishable from the bulk $T_S$.

Figure \ref{fig:decayg2} summarizes the three possible phase-transition scenarios with the lattice distortion effects: (i) XY-alternating order with surface instability for $\bar{g}<\bar{g}_{c1}$, (ii) uniform order with surface instability for $\bar{g}_{c1}<\bar{g}<\bar{g}_{c2}$, and (iii) uniform bulk order without surface instability for $\bar{g}_{c2}<\bar{g}$. 
We explored in detail these different transition scenarios for $\bar{t}_z=0.05$ and $\bar{t}'_z=0$ and the results are summarized in the phase diagram presented in Fig.\ \ref{fig:SI}. The most interesting scenario, uniform bulk order with surface instability, is realized within the intermediate range of $\bar{g}$ highlighted by the shaded area in the phase diagram. We can see that this range rapidly grows with increasing the ratio $C_s/C'_s$ and can be rather wide.


\section{Discussion and Conclusions}\label{finale}

To summarize our work, we have considered the bulk and surface nematic phase transitions in layered materials. The consideration is based on the simple single-layer two-band mean-field Hamiltonian and electronic tunneling between the layers. Evaluating the nematic free energy near the critical point, we have demonstrated that nematic order forms near surface before the main bulk transition. We also found that purely electronic tunneling Hamiltonian favors XY-alternating order in which sign of the order parameter changes from layer to layer.

Furthermore, lattice elasticity plays an important role in determining the ordering pattern in the multilayer system. In particular, coupling to the lattice may stabilize the uniform bulk nematic order.  Depending on the coupling strength between the electrons and lattice, we found three different scenarios: (i) XY-alternating order with surface transition, (ii) Uniform order with surface transition, and (iii) Uniform bulk transition without surface instability. Scenario (ii) spans a considerable region in the parameter space and may 
plausible for iron-pnictides. The surface nematic may be realization of the intermediate nematic state reported for P-doped 122 materials.\cite{Kasahara:2012fk,2015arXiv150703981T} As several powerful experimental techniques, such as STM and ARPES, are inherently surface probes, the formation of preemptive surface nematic may strongly influence interpretation of experimental data.

Typically nematic instability in iron pnictides occurs either simultaneously or in the close proximity with the antiferromagnetic transition
meaning that the spin fluctuations may strongly influence the formation of the nematic order.
A proper microscopic treatment of these fluctuations is complicated and requires consideration of the hole bands in the zone center, which we leave for the future work.   On the phenomenological level,  the nematic order couples to the spin fluctuations linearly in the Landau-theory free-energy expansion, \cite{Fernandes:2014sf} 
$\delta F \propto \Delta_\ell (M^2_{X,\ell}-M^2_{Y,\ell})$, where $M_{\alpha,\ell}$ are the fluctuating stripe-antiferromagnetic magnetizations in two perpendicular directions.
Integrating out the spin fluctuations, this linear-coupling term generates the negative correction to the inverse nematic susceptibility which decays with $|\ell-\ell'|$. This means that the spin fluctuations promote the nematic ordering in each layer and favor the uniform order with respect to the XY-alternating order.
We also expect that these fluctuations should suppress the surface instability. Therefore, effects of the spin fluctuations are qualitatively similar to ones of coupling to the lattice distortions.   

In our study, the e-e interactions between different layers are ignored by assuming that overlapping between the Fe-orbital wave functions is negligible, since FeSC are layered materials. If the interlayer e-e interactions play an important role in driving the nematic order, long-range correlation can be built up between different layers. In this case, bulk nematic order can be more energetically favorable. 
We also note that the approach in this paper is only valid for the study of second-order phase transition near the critical temperature. To accurately describe the case of first-order phase transition or far away from the critical temperature, higher-order terms in the free energy have to be taken into account. These higher-order terms may change the ground state configuration drastically.

\begin{acknowledgements}
We would like to thank Ian Fisher for useful discussion. This work was supported by the Center for Emergent Superconductivity, an Energy Frontier Research Center funded by the US DOE, Office of Science, under Award No. DEAC0298CH1088.
\end{acknowledgements}


\appendix

\section{Interlayer tight-binding model and $XY$-FS pocket hybridization}
\label{app:HXY}

In iron pnictides with 122 composition, such as BaFe$_2$As$_2$, the off-diagonal out-of-plane hopping is important, because it modifies electronic spectrum qualitatively. The tight-binding Hamiltonian that includes the off-diagonal hopping in term of orbital basis is 
\begin{equation}\label{Hhyb1}
\begin{split}
\mathcal{H}_{\mathrm{tun}}=-&\Big[\sum^{\text{odd}}_{\ell}\Big(\sum_{\mathbf{n}=\mathbf{n}_1}\sum_{\delta=\delta_1}+\sum_{\mathbf{n}=\mathbf{n}_2}\sum_{\delta=\delta_2}\Big)+\\
&\sum^{\text{even}}_{\ell}\Big(\sum_{\mathbf{n}=\mathbf{n}_1}\sum_{\delta=\delta_2}+\sum_{\mathbf{n}=\mathbf{n}_2}\sum_{\delta=\delta_1}\Big)\Big] \times\\
&\sum_{\bar{o}\bar{o}'}h^{\bar{o}\bar{o}'}d^\dagger_{\bar{o},\ell,\mathbf{n}s}d_{\bar{o}',\ell+1,\mathbf{n}+\delta,s}+h.c.,
\end{split}
\end{equation}
where $d^\dagger_{\bar{o},\ell,\mathbf{n}}$ ($d_{\bar{o},\ell,\mathbf{n}}$) is the orbital creation (annihilation) field operator with orbital index $\bar{o}=1,2,3$ standing for $d_{xz}$, $d_{yz}$, and $d_{xy}$ respectively. Furthermore, $h^{\bar{o}\bar{o}'}$ is the tight-binding constant, 
$\ell$ is the layer index, and $\mathbf{n}=(n_x,n_y)$ is the lattice site index in the Fe-layer, and $\mathbf{n}_1$ ($\mathbf{n}_2$) is the lattice site with $n_{x}+n_{y}=\text{odd}$ ($n_{x}+n_{y}=\text{even}$). The following next-nearest hoppings have almost equal strength $\delta_1=(0,0), (\pm 1,0), (0,\pm 1), (-1,1), (1,-1)$  and $\delta_2=(0,0),(\pm 1,0), (0,\pm 1), (-1,-1), (1,1)$.

We express the orbital field operators in the momentum space as
\begin{equation}
d_{\bar{o},\ell,\mathbf{n}s}=\frac{1}{\sqrt{A}}\sum_{\mathbf{k}}e^{-i\mathbf{k}\cdot\mathbf{n}}d_{\bar{o},\ell,\mathbf{k}s},
\end{equation}
where $A$ is the total number of unit-cells in the Fe-layer. Substituting the above equation into $\mathcal{H}_{\mathrm{tun}}$, we break the tight-binding Hamiltonian as follows: $\mathcal{H}_{\mathrm{tun}}=\mathcal{H}^0_{\mathrm{tun}}+\mathcal{H}'_{\mathrm{tun}}+\mathcal{H}_{\mathrm{hyb}}$. 

For the direct hopping: $\delta_1=\delta_2=(0,0)$,
\begin{equation}
\mathcal{H}^0_{\mathrm{tun}}=\sum^{N-1}_{\ell}\sum_{\bar{o}\bar{o}',\mathbf{k}}h^{\bar{o}\bar{o}'}d^\dagger_{\bar{o},\ell,\mathbf{k}s}d_{\bar{o}',\ell+1,\mathbf{k}s}+h.c.
\end{equation}
Note that, in this case, we have combined the even and odd layers and sub-lattice in the summation of $\ell$ and $\mathbf{n}$. 
The  orbital field operator 
can be expressed in term of band electron field operator as follows. 
\begin{align}
d_{\bar{o},\ell,\mathbf{k}s}&=\sum_{\alpha}[\gamma^{\bar{o}}_{\bar{\alpha}\mathbf{k}}]^{-1} c_{\ell,\bar{\alpha},\mathbf{k}s}.\label{bandbasis}
\end{align}
where $\gamma^{\bar{o}}_{\bar{\alpha}\mathbf{k}}$ is the rotation matrix which diagonalized the single layer tight-binding Hamiltonian in the $\mathbf{k}$-space, and $\bar{\alpha}$ is the corresponding band index.
Substituting the  
Eq. \eqref{bandbasis}
into Eq. \eqref{Hhyb1}, we obtain
\begin{equation*}
\mathcal{H}^0_{\mathrm{tun}}=-\sum^{N-1}_{\ell}\sum_{\mathbf{k};\bar{\alpha}\bar{\alpha}'}\lambda^{\bar{\alpha}\bar{\alpha}'}_{1,\mathbf{k}}c^\dagger_{\ell,\bar{\alpha},\mathbf{k}s}c_{\ell+1,\bar{\alpha}',\mathbf{k}s}+h.c.,
\end{equation*}
where $\lambda^{\bar{\alpha}\bar{\alpha}'}_{1,\mathbf{k}}=\sum_{\bar{o}\bar{o}'}h^{\bar{o}\bar{o}'}[(\gamma^{\bar{o}}_{\bar{\alpha}\mathbf{k}})^\ast]^{-1} [\gamma^{\bar{o}'}_{\bar{\alpha}\mathbf{k}}]^{-1}$. If we ignore the $\mathbf{k}$ dependence in $\lambda_1$ and restrict the momentum near the FS then $\lambda^{\bar{\alpha}\bar{\alpha}'}_{1,\mathbf{k}}\simeq t_z$ which is the direct hopping term in the Hamiltonian \eqref{HN}.

For the next-nearest neighbor hoppings: $\delta_1= (\pm 1,0), (0,\pm 1)$,
\begin{equation}
\begin{split}
\mathcal{H}'_{\mathrm{tun}}=-\sum^{N-1}_{\ell}\sum_{\mathbf{k};\bar{\alpha}\bar{\alpha}'}2\lambda^{\bar{\alpha}\bar{\alpha}'}_{1,\mathbf{k}}(\cos k_x+\cos k_y )\times\\
(c^\dagger_{\ell,\bar{\alpha},\mathbf{k}s}c_{\ell+1,\bar{\alpha}',\mathbf{k}s}+h.c.).
\end{split}
\end{equation}
This term modifies the interlayer nearest-neighbor hopping constant. 

Turning to the next-next nearest-neighbor hopping: $\delta_1=(1,-1), (-1,1)$ and $\delta_2=(1,1), (-1,-1)$, we derive
\begin{widetext}
\begin{equation}
\begin{split}
\mathcal{H}_{\mathrm{hyb}}=&-\frac{1}{A}\sum^\text{odd}_{\ell}\sum_{\bar{o}\bar{o}'}h^{\bar{o}\bar{o}'}\sum_{\mathbf{k}\mathbf{k}'}\sum_{\mathbf{n}}2(\cos k'_x\cos k'_y -e^{i\mathbf{Q}\cdot\mathbf{n}}\sin k'_x\sin k'_y)e^{i(\mathbf{k}-\mathbf{k}')\cdot\mathbf{n}}d^\dagger_{\bar{o},\ell,\mathbf{k}s}d_{\bar{o}',\ell+1,\mathbf{k}'s}+h.c.\\
&-\frac{1}{A}\sum^\text{even}_{\ell}\sum_{\bar{o}\bar{o}'}h^{\bar{o}\bar{o}'}\sum_{\mathbf{k}\mathbf{k}'}\sum_{\mathbf{n}}2(\cos k'_x\cos k'_y +e^{i\mathbf{Q}\cdot\mathbf{n}}\sin k'_x\sin k'_y)e^{i(\mathbf{k}-\mathbf{k}')\cdot\mathbf{n}}d^\dagger_{\bar{o},\ell,\mathbf{k}s}d_{\bar{o}',\ell+1,\mathbf{k}'s}+h.c.
\end{split}
\end{equation}
Note that in the calculation we have used $(-1)^{n_x+n_y}=e^{i\mathbf{Q}\cdot\mathbf{n}}$.  
Carrying out the $\mathbf{n}$ summation and combining the $\ell$ = even and odd terms, we obtain
\begin{equation}\label{H_hybpp}
\mathcal{H}_{\mathrm{hyb}}=-\sum^{N-1}_{\ell}\sum_{\bar{o}\bar{o}'}h^{\bar{o}\bar{o}'}\sum_{\mathbf{k}}2(\cos k_x\cos k_y d^\dagger_{\bar{o},\ell,\mathbf{k}s}d_{\bar{o}',\ell+1,\mathbf{k}s} +(-1)^\ell\sin k_x\sin k_yd^\dagger_{\bar{o},\ell,\mathbf{k}s}d_{\bar{o}',\ell+1,\mathbf{k}+\mathbf{Q},s})+h.c.
\end{equation}
\end{widetext}
The momentum $\mathbf{k}$ is measured from $(0,\pi)$ and $\mathbf{Q}=(\pi,\pi)$. The last term generates the hybridization between $X$- and $Y$- pockets. 
To see this, we write $\mathcal{H}''_{\mathrm{hyb}}$ in the band basis, and keeping only the last term in Eq. \eqref{H_hybpp},
\begin{equation}
\begin{split}
\mathcal{H}_{\mathrm{hyb}}=\sum_\ell^{N-1}\sum_{\mathbf{k},\bar{\alpha}\bar{\alpha}'}(-1)^{\ell+1}2\lambda^{\bar{\alpha}\bar{\alpha}'}_{2,\mathbf{k}}\sin k_x \sin k_y \\ \times c^\dagger_{\bar{\alpha},\ell,\mathbf{k}s}c_{\bar{\alpha}',\ell+1,\mathbf{k}+\mathbf{Q},s},
\end{split}
\end{equation}
where $\lambda^{\bar{\alpha}\bar{\alpha}'}_{2,\mathbf{k}}=\sum_{\bar{o}\bar{o}'}h^{\bar{o}\bar{o}'}[(\gamma^{\bar{o}}_{\bar{\alpha}\mathbf{k}})^\ast]^{-1} [\gamma^{\bar{o}'}_{\bar{\alpha}\mathbf{k}+\mathbf{Q}}]^{-1}$. If we restrict the momentum to be near the FS, and regroup the band index into $X$ and $Y$ according to their momentum, this immediately lead to the $X$- and $Y$- pockets hybridization. As in the direct hopping term, for simplicity, ignoring the $\mathbf{k}$-dependence in $\lambda^{\bar{\alpha}\bar{\alpha}'}_{2,\mathbf{k}}$ and set it to $t_z'$, we therefore obtain 
\begin{equation}
\begin{split}
\mathcal{H}_{\mathrm{hyb}}\simeq\sum^{N-1}_{\ell=1}\sum_\mathbf{k}\lambda_\mathbf{k}(-1)^{\ell-1}(c^\dagger_{X,\ell,\mathbf{k}s}c_{Y,\ell+1,\mathbf{k},s}
\\+c^\dagger_{Y,\ell,\mathbf{k}s}c_{X,\ell+1,\mathbf{k},s})+h.c.
\end{split}
\end{equation}
with $\lambda_\mathbf{k}=2t'_z\sin k_x\sin k_y$.
\vspace{0.15in}
\section{Derivation of the bulk free energy}\label{app:Hinfty}

In this section, we derive the free energy in $N\to\infty$ limit. 
By the definition of free energy,
\begin{equation}
F'_b[\tilde{\Delta}_q]\!=\!\int_{q}\frac{\tilde{\Delta}_q\tilde{\Delta}_{-q}}{2u}-\frac{2}{\beta S}\text{tr}\ln[i\omega_n\!-\!\hat{H}'_b\!-\!\hat{V}'_b\!+\!\mu].
\end{equation}
To obtain the inverse nematic susceptibility, we expand the free energy up to second order in $\tilde{\Delta}$,
\begin{align*}
F^{(2)}_b\!=\!\int_q\frac{|\tilde{\Delta}_q|^2}{2 u}
\!+\!\frac{1}{\beta}\sum_{\substack{k_z,k^{\prime}_z\\\omega_n}}\int_\mathbf{k}\text{tr}(\mathscr{G}'_{k_z}\mathscr{V}'_{k_z-k'_z}\mathscr{G}'_{k'_z}\mathscr{V}'_{k'_z-k_z}),
\end{align*}
where $\mathscr{V}'_{k_z-k'_z}$ is given by Eq. \eqref{Vb}, and 
\begin{widetext}
\[\mathscr{G}_{k_z}=\begin{pmatrix}
[(i\omega_n-\varepsilon^Y_\mathbf{k}+\mu)\mathbb{I}+2t_\mathbf{k}\sigma^z\cos k_z]\Omega^{-1}_{k_z} & -2\lambda_\mathbf{k}\sigma^y\sin k_z\Omega^{-1}_{k_z+\pi} \\
-2\lambda_\mathbf{k}\sigma^y \sin k_z\Omega^{-1}_{k_z} & [(i\omega_n-\varepsilon^X_\mathbf{k}+\mu)\mathbb{I} +2t_\mathbf{k}\sigma^z\cos k_z]\Omega^{-1}_{k_z+\pi}
\end{pmatrix}
\]
with
\[
\Omega_{k_z}=\begin{pmatrix}(G^X_{k_z}G^Y_{k_z+\pi})^{-1}-4\lambda_\mathbf{k}^2 \sin^2 k_z&0\\
0&(G^X_{k_z+\pi}G^Y_{k_z})^{-1}-4\lambda_\mathbf{k}^2 \sin^2 k_z
\end{pmatrix}
\]
and $G^\alpha_{k_z}=i\omega_n-\varepsilon^\alpha_\mathbf{k}+2t_\mathbf{k}\cos k_z+\mu$. Writing out the trace explicitly and using the periodic condition in $q\to q+2\pi$, this yields
\begin{equation*}
F^{(2)}_b=
-\frac{1}{2}\int_{q}r'_{b,q}\tilde{\Delta}_{-q}\tilde{\Delta}_{q}
\end{equation*}
with the inverse nematic susceptibility 
\begin{equation*}
\begin{split}
r'_{b,q}=&\frac{1}{u}-2\int_{\mathbf{k}k_z}\text{Res}\Big[
	\frac{[(z-z^Y_{k_z+\pi})(z-z^Y_{k_z+q+\pi})-(2\lambda_\mathbf{k})^2\sin k_z\sin (k_z+q)]\frac{1}{2}\tanh\frac{\beta z}{2}}
		{[(z-z^X_{k_z})(z-z^Y_{k_z+\pi})
		-4\lambda^2_\mathbf{k}\sin^2 k_z][(z-z^X_{k_z+q})(z-z^Y_{k_z+q+\pi})
		-4\lambda^2_\mathbf{k}\sin^2 (k_z+q)]}\Big]\\
		&+(X\leftrightarrow Y)
\end{split}
\end{equation*}
where $z^\alpha_{k_z}=\varepsilon^\alpha_\mathbf{k}-\mu-2t_\mathbf{k}\cos k_z$. Note that the summation of $k_z$ is running over $k_z\in[-\pi,\pi]$.
The denominator has poles at $z=z^{\pm}_{k_z}=\varepsilon^{\pm}_{\mathbf{k}k_z}-\mu$ and $z=z^{\pm}_{k_z+q}=\varepsilon^{\pm}_{\mathbf{k},k_z+q}-\mu$, where $\varepsilon^{\pm}_{\mathbf{k}k_z}$ is defined in Eq. \eqref{eqn:epm}.
Factorizing the denominator with these poles, and using the symmetry by exchanging $X\leftrightarrow Y$, we obtain
\begin{equation}\label{Res_of_susinfty}
r'_{b,q}=
\frac{1}{u}-2\int_{\mathbf{k}k_z}
	\text{Res}\Big[
	\frac{[(z-z^Y_{k_z+\pi})(z-z^Y_{k_z+q+\pi})-(2\lambda_\mathbf{k})^2\sin (k_z+\pi)\sin (k_z+q+\pi)]\tanh\frac{\beta z}{2}}
		{(z-z^+_{k_z})(z-z^-_{k_z})(z-z^+_{k_z+q})(z-z^-_{k_z+q})}
	\Big]
\end{equation}
Applying the residue theorem straightforwardly, we finally obtain Eq.\ \eqref{rbqGen}.

For the rest of this section, we evaluate the inverse susceptibility for some special cases. For circular, FS $\delta_\mathbf{k}=0$, the susceptibility reduced to
	\begin{align}
		\label{rbq}
	&r'_{b,q}=\frac{1}{u}
	-\int_{\mathbf{k},k_z}\frac{1}{4(t_\mathbf{k}^2-\lambda^2_\mathbf{k})}
	\left[
	\frac{[\eta^2_{\mathbf{k}k_z}+4t^2_\mathbf{k}\cos k_z\cos(k_z+q)-4\lambda^2_\mathbf{k}\sin k_z\sin(k_z+q)]n_-(k_z)}
	{\eta_{\mathbf{k}k_z}(\cos^2 (k_z+q)-\cos^2 k_z)}\right.\notag\\
	&\left.
	-\frac{[\eta^2_{\mathbf{k},k_z+q}+4t^2_\mathbf{k}\cos k_z\cos(k_z+q)-4\lambda^2_\mathbf{k}\sin k_z\sin(k_z+q)]n_-(k_z+q)}
	{\eta_{\mathbf{k},k_z+q}(\cos^2 (k_z+q)-\cos^2 k_z)}
	-\frac{2t_\mathbf{k}[n_+(k_z)-n_+(k_z+q)]}
	{\cos (k_z+q)-\cos k_z}
	\right]
	\end{align}
where $n_{\pm}(k_z)=\tanh(\beta z^+_{k_z}/2)\pm\tanh(\beta z^-_{k_z}/2)$, $z^\pm_{k_z}=\mathbf{k}^2/(2m)-\mu\pm\eta_{\mathbf{k}k_z}$, and $\eta_{\mathbf{k}k_z}=2(t^2_\mathbf{k}\cos^2 k_z+\lambda^2_\mathbf{k}\sin^2k_z)^{1/2}$.
Furthermore, using equation \eqref{Res_of_susinfty} with $\delta_\mathbf{k}=0$, we have
\begin{equation}\label{rb0}
\begin{split}
r'_{b,q=0}=\frac{1}{u}
-2\int_{\mathbf{k}k_z}
	\Big[&
	\frac{2\lambda^2_\mathbf{k}\sin^2 k_z\left(\tanh\frac{\beta z^+_{k_z}}{2}-\tanh\frac{\beta z^-_{k_z}}{2}\right)}{\eta_{\mathbf{k}k_z}^3}
	+\frac{\beta t^2_\mathbf{k}\cos^2 k_z}{\eta_{\mathbf{k},k_z}^2}\left(\mathrm{sech}^2\tfrac{\beta z^+_{k_z}}{2}+\mathrm{sech}^2\tfrac{\beta z^-_{k_z}}{2}\right)\\
	&-\frac{t_\mathbf{k}\cos k_z}{\eta_{\mathbf{k},k_z}}\frac{\beta}{2}\left(\mathrm{sech}^2\tfrac{\beta z^+_{k_z}}{2}-\mathrm{sech}^2\tfrac{\beta z^-_{k_z}}{2}\right)\Big]
\end{split}
\end{equation}
Similarly, for $q=\pi$ with $\delta_\mathbf{k}=0$,
\begin{equation}\label{rbpi}
r'_{b,q=\pi}=\frac{1}{u}
-2\int_{\mathbf{k}k_z}
	\Big[\frac{2t^2_\mathbf{k}\cos^2 k_z}{\eta_{\mathbf{k}k_z}^3}\left(\tanh\tfrac{\beta z^+_{k_z}}{2}-\tanh\tfrac{\beta z^-_{k_z}}{2}\right)
	+\frac{\beta\lambda_\mathbf{k}\sin^2 k_z}{\eta_{\mathbf{k}k_z}^2}\left(\mathrm{sech}^2\tfrac{\beta z^+_{k_z}}{2}+\mathrm{sech}^2\tfrac{\beta z^-_{k_z}}{2}\right)\Big]
\end{equation}
\end{widetext}

\section{Tight-binding Green's function $\mathcal{G}^{\alpha}_{0}$}\label{app:Ginverse}

To find $\mathcal{G}^{\alpha}_{0}$ in section \ref{NN}, we first note that the eigenmodes of the Hamiltonian operator without hybridization is 
\begin{equation}\label{pbasis}
c_{\alpha,p,\mathbf{k}s}=\sqrt{\frac{2}{N+1}}\sum_{\ell}\sin \ell\vartheta_p c_{\alpha,\ell,\mathbf{k}s},
\end{equation}
where $\vartheta_p=\frac{p\pi}{N+1}$ with $p=1\dots N$. This can be checked by substituting the above equation into the Hamiltonian
\begin{equation}
\begin{split}
\mathcal{H}^0_{N}=&\sum^{N}_{\ell=1}\sum_{\alpha,\mathbf{k}}\varepsilon^\alpha_{\mathbf{k}}c^\dagger_{\alpha,\ell,\mathbf{k}\sigma}c_{\alpha,\ell,\mathbf{k}\sigma}\\
&-t_z\sum^{N-1}_{\ell=1}\sum_{\alpha,\mathbf{k}}c^\dagger_{\alpha,\ell,\mathbf{k}\sigma}c_{\alpha,\ell+1,\mathbf{k}\sigma}+h.c.
\end{split}
\end{equation}
By using the orthogonal relation $\sum_{\ell}\sin \ell\vartheta_p\sin \ell\vartheta_{p'}=\frac{2}{N+1}\delta_{pp'}$, we obtain
\begin{equation}
\begin{split}
\mathcal{H}^0_{N}=\sum^{N}_{p=1}\sum_{\alpha,\mathbf{k}}(\varepsilon^\alpha_{\mathbf{k}}-2t_z\cos\vartheta_p)c^\dagger_{\alpha,p,\mathbf{k}\sigma}c_{\alpha,p,\mathbf{k}\sigma},
\end{split}
\end{equation}
which is diagonal in `$p$'-basis. 

Therefore, expanding the Green's function in this basis, this immediately lead to
\begin{equation}
[\mathcal{G}^{\alpha}_{0}]_{\ell\ell'}=\sum_{p}\frac{2/(N+1)\sin \ell\vartheta_p\sin \ell'\vartheta_p}{i\omega_n-\varepsilon^\alpha_{\mathbf{k}}+2t_z\cos\vartheta_p+\mu}.
\end{equation}

\section{Small-$t_z$ expansion of the inverse susceptibility $r_{\ell\ell'}$ for finite-size system with $t'_z=0$}\label{smalltz}

To make the expansion with respect to $t_z$, we start from the following presentation for the second term in Eq.\ \eqref{rll0}
\begin{equation}\label{Resrll}
\frac{2}{\beta}\sum_{\omega_n}[\mathcal{G}^\alpha_0]_{\ell\ell'}[\mathcal{G}^\alpha_0]_{\ell'\ell}\!=\!-\!\sum_{pp'}\text{Res}\left[\frac{S^{p}_{\ell\ell'}S^{p'}_{\ell'\ell}\tanh\frac{\beta z}{2}}{(z\!-\!z^\alpha_p)(z\!-\!z^\alpha_{p'})}\right],
\end{equation}
where the analytic-continuation technique has been used in the frequency summation, and $z^\alpha_p=\xi_{\alpha,\mathbf{k}}-2t_z\cos\vartheta_p$ with $\alpha\!=\!X,Y$, $\xi_{X,\mathbf{k}}=k_x^2/(2m_x)+k_y^2/(2m_y)-\mu$, $\xi_{Y,\mathbf{k}}=k_x^2/(2m_y)+k_y^2/(2m_x)-\mu$, and $S^{p}_{\ell\ell'}=2\sin(\ell\vartheta_p)\sin(\ell'\vartheta_{p})/(N\!+\!1)$.
Using the orthogonality relation $\sum^N_{p=1}\sin(\ell\vartheta_p)\sin(\ell'\vartheta_{p})\!=\!\delta_{\ell\ell'}(N\!+\!1)/2$, the expansion of Eq. \eqref{Resrll} is
\begin{equation*}
\begin{split}
\text{Res}\left[\frac{\tanh\frac{\beta z}{2}}{(z-\xi_\mathbf{k})^2}\Big(\delta_{\ell'\ell}+\frac{2t_z\delta_{\ell\ell'}(\delta_{\ell,\ell'+1}+\delta_{\ell,\ell'-1})}{(z-\xi_\mathbf{k})}\right.\\
\left.+t^2_z\frac{(\delta_{\ell,\ell'+1}+\delta_{\ell,\ell'-1})^2}{(z-\xi_\mathbf{k})^2}\right.
\left.+4t^2_z\frac{(1-\frac{\delta_{\ell,1}+\delta_{\ell,N}}{2})\delta_{\ell\ell'}}{(z-\xi_\mathbf{k})^2}\Big)
\right],
\end{split}
\end{equation*}
where we used
\begin{align*}
&\sum_{p}^NS_{\ell\ell'}^p\cos\vartheta_p=\frac{\delta_{\ell,\ell'+1}+\delta_{\ell,\ell'-1}}{2},\\
&\sum_{p}^NS_{\ell\ell}^p\cos^2\vartheta_p=1-\frac{\delta_{\ell,1}+\delta_{\ell,N}}{2}.
\end{align*}
Applying the residue theorem, we immediately obtain the approximation of $r_{\ell\ell'}$ for small $t_z$,
\begin{equation}
\begin{split}
r_{\ell\ell'}&\simeq
\left[\frac{1}{u}-\beta\int_\mathbf{k}\mathrm{sech}^2\tfrac{\beta \xi_\mathbf{k}}{2}\right]\delta_{\ell\ell'}\\
&-\frac{\beta^3t^2_z}{3}\int_\mathbf{k}\Big[\mathrm{sech}^2\tfrac{\beta \xi_\mathbf{k}}{2}(3\tanh^2\tfrac{\beta \xi_\mathbf{k}}{2}-1)\Big]\times\\
&\Big[\delta_{\ell\ell'}-\frac{\delta_{\ell,1}+\delta_{\ell,N}}{2}\delta_{\ell\ell'}+\frac{\delta_{\ell,\ell'+1}+\delta_{\ell,\ell'-1}}{4}\Big]
\end{split}
\end{equation}
Integration over the in-plane momentum $\mathbf{k}$ using $\int_\mathbf{k}\rightarrow (\tilde{m}/2\pi)\int d\xi_\mathbf{k}$ gives the result \eqref{eq:smalltz} of the main text. 

\section{Perturbation calculation for the case when $N$ is odd and $m_x=m_y$}\label{degen}

For $N$ is odd with isotropic FS ($m_x=m_y$), the matrix in the first term of Eq.\ \eqref{Hhat} is degenerate. In this case, Eq.\ \eqref{2ndpert} cannot be used directly due to zero denominators in some terms. 
Before computing the eigenvalue of $\hat{H}$ pertubatively,
we first rotate the eigenspace of $\hat{H}$ matrix in $F'_N[\Delta_\ell]$ by 
\[
R=\frac{1}{\sqrt{2}}\begin{pmatrix}
\mathbb{I} & -\mathbb{I}\\
\mathbb{I} & \mathbb{I}
\end{pmatrix}
\]
where $\mathbb{I}$ is a $N\times N$ identity matrix. Namely,
\begin{equation*}
\begin{split}
F'_N[\Delta_\ell]&=\sum_{\ell}\frac{\Delta^2_{\ell}}{2u}+\frac{2}{S \beta}\text{tr}[R^{-1}\ln(i\omega_n -\hat{H})R]\\
&=\sum_{\ell}\frac{\Delta^2_{\ell}}{2u}+\frac{2}{S\beta}\text{tr}[\ln(i\omega_n -R^{-1}\hat{H}R)].
\end{split}
\end{equation*}
Therefore, we obtain
\begin{equation*}
R^{-1}\hat{H}R=
\begin{pmatrix}
\mathbf{E}_\mathbf{k}+\mathbf{P}_\mathbf{k}&0\\
0&\mathbf{E}_\mathbf{k}-\mathbf{P}_\mathbf{k}
\end{pmatrix}+
\begin{pmatrix}
0&-\mathcal{V}\\
-\mathcal{V}&0
\end{pmatrix},
\end{equation*}
where $\mathbf{E}_\mathbf{k}=\mathbf{E}^X_\mathbf{k}=\mathbf{E}^Y_\mathbf{k}$ is given by equation \eqref{Ek} (the $X$ and $Y$ pockets are identical), $\mathbf{P}_\mathbf{k}$ is defined by Eq.\ \eqref{Pk}, and $[\mathcal{V}]_{\ell\ell'}=\Delta_\ell\delta_{\ell\ell'}$. The first term in $R^{-1}\hat{H}R$ becomes block-diagonalized and more convenient for perturbation calculation. The second term in $R^{-1}\hat{H}R$ is treated as perturbation.

Now, we let $\tilde{\mathbf{x}}_{p,\mathbf{k}}^T=[\tilde{x}^1_{p,\mathbf{k}},\dots \tilde{x}^N_{p,\mathbf{k}}]$ and $\tilde{\mathbf{y}}_{p,\mathbf{k}}^T=[\tilde{y}^1_{p,\mathbf{k}},\dots \tilde{y}^N_{p,\mathbf{k}}]$ with $p=1\dots N$, and they satisfies
\begin{equation*}
\begin{split}
(\mathbf{E}_\mathbf{k}+\mathbf{P}_\mathbf{k})\tilde{\mathbf{x}}_{p,\mathbf{k}}=f^{(0)}_{p,\mathbf{k}}\tilde{\mathbf{x}}_{p,\mathbf{k}},\\
(\mathbf{E}_\mathbf{k}-\mathbf{P}_\mathbf{k})\tilde{\mathbf{y}}_{p,\mathbf{k}}=f^{(0)}_{p,\mathbf{k}}\tilde{\mathbf{y}}_{p,\mathbf{k}},
\end{split}
\end{equation*}
where $f^{(0)}_{p,\mathbf{k}}$ is the unperturbed eigenvalue.

Furthermore, in order to make connection with the standard notation in quantum mechanics perturbation theory, we introduce the following `bra' and `ket' notation.
\begin{align}
|\tilde{\mathbf{x}}_p \rangle=\begin{pmatrix}
\tilde{\mathbf{x}}_{p,\mathbf{k}} \\ 
\mathbf{0}
\end{pmatrix},\quad
|\tilde{\mathbf{y}}_p \rangle=\begin{pmatrix}
\mathbf{0}\\
\tilde{\mathbf{y}}_{p,\mathbf{k}}
\end{pmatrix},
\end{align}
where $\mathbf{0}$ is a $1\times N$ zero matrix. Thus, $|\tilde{\mathbf{x}}_p \rangle$ and $|\tilde{\mathbf{y}}_p \rangle$ span the $p$-th 2-fold degenerate subspace. Also, we set the perturbation operator as
\begin{equation}
\hat{V}=\begin{pmatrix}
0&-\mathcal{V}\\
-\mathcal{V}&0
\end{pmatrix}.
\end{equation}
One can immediately see that, any linear combination of $|\tilde{\mathbf{x}_p}\rangle$ and $|\tilde{\mathbf{y}_p}\rangle$ are still the eigenvector of the first term in $R^{-1}\hat{H}R$. Therefore, the choices of eigenvector are not unique. Exploiting this fact, we can choose a basis such that the numerators with  overlapping degenerate eigenvectors in Eq. \eqref{2ndpert} vanish. Hence, the zero denominator terms are dropped out in the calculation. The procedure to obtain such basis is as follows.

First, we calculate the first-order correction for the eigenvalue in the $p$-th degenerate subspace. In this subspace, the operator $\hat{V}$ can be represented as the following matrix form, 
\begin{equation*}
\begin{pmatrix}
\langle\tilde{\mathbf{x}}_p|\hat{V}|\tilde{\mathbf{x}}_p \rangle & \langle\tilde{\mathbf{x}}_p|\hat{V}|\tilde{\mathbf{y}}_p \rangle\\
\langle\tilde{\mathbf{y}}_p|\hat{V}|\tilde{\mathbf{x}}_p \rangle&\langle\tilde{\mathbf{y}}_p|\hat{V}|\tilde{\mathbf{y}}_p \rangle
\end{pmatrix}
=\begin{pmatrix}
0 & \tilde{\mathbf{x}}_{p,\mathbf{k}}^T\mathcal{V}\tilde{\mathbf{y}}_{p,\mathbf{k}}\\
\tilde{\mathbf{y}}_{p,\mathbf{k}}^T\mathcal{V}\tilde{\mathbf{x}}_{p,\mathbf{k}} & 0
\end{pmatrix}.
\end{equation*}
Solving the eigenvalues of the above $2\times2$ matrix yields the first-order correction. Writing out the nematic order parameters explicitly, this matrix becomes
\begin{equation}
\sum_\ell\begin{pmatrix}0& \tilde{x}^\ell_{p,\mathbf{k}}\tilde{y}^\ell_{p,\mathbf{k}}\Delta_\ell\\
 \tilde{x}^\ell_{p,\mathbf{k}}\tilde{y}^\ell_{p,\mathbf{k}}\Delta_\ell&0\end{pmatrix},
\end{equation}
and has the following eigenvalues and eigenvectors 
\begin{equation}\label{xyv}
\pm \sum_\ell \tilde{x}^\ell_{p,\mathbf{k}}\tilde{y}^\ell_{p,\mathbf{k}}\Delta_\ell \text{, and } \frac{1}{\sqrt{2}}\begin{pmatrix}1\\ \pm1\end{pmatrix}.
\end{equation}
These results yield the desirable `rotated' eigenvectors \begin{equation}\label{ppm}
|p,\pm\rangle=\frac{1}{\sqrt{2}}(|\tilde{\mathbf{x}}_p \rangle\pm|\tilde{\mathbf{y}}_p \rangle)
\end{equation} 
with the first-order corrected eigenvalue
\begin{equation}
\tilde{f}^{(p,\pm)}_\mathbf{k}\simeq \tilde{f}^{(0)}_{p,\mathbf{k}}\pm\sum_{\ell}\tilde{a}^{(p,\pm)}_\ell\Delta_\ell+\mathcal{O}(\Delta^2),
\end{equation}
where $\tilde{a}^{(p,\pm)}_{\ell}=\pm\tilde{x}^\ell_{p,\mathbf{k}}\tilde{y}^\ell_{p,\mathbf{k}}$. The first order correction has lifted the degeneracy and single out the particular choice of linear combination: $|p,\pm\rangle$. Also, note that, only this choice can be smoothly approached from the perturbed eigenvectors when the perturbations are turning off.

Further, the second-order corrected eigenvalues for eigenvectors $|p, \pm\rangle$, Eq.\ \eqref{ppm}, are evaluated as 
\begin{equation}\label{2ndcorr}
\begin{split}
&\sum_{p'\neq p}\frac{|\langle p,\pm|\hat{V}|p',+\rangle|^2+|\langle p,\pm|\hat{V}|p',-\rangle|^2}{f^{(0)}_{p,\mathbf{k}}-f^{(0)}_{p',\mathbf{k}}}.
\end{split}
\end{equation}
Note that, in the summation, not only the terms $\langle p,\pm|\hat{V}|p,\pm\rangle$ are excluded, but also $\langle p,-|\hat{V}|p,+\rangle$ and $\langle p,+|\hat{V}|p,-\rangle$ (also having zero denominator), since they vanish in the rotated new basis.

Therefore, the approximation of the $2N$ eigenvalues up to second order is
\begin{equation}\label{tildefapprox}
\tilde{f}_{p,\pm,\mathbf{k}}\simeq \tilde{f}^{(0)}_{p,\mathbf{k}}+\sum_\ell \tilde{a}^{p,\pm}_\ell\Delta_\ell+\sum_{\ell\ell'}\tilde{b}^{p,\pm}_{\ell\ell'}\Delta_\ell\Delta_{\ell'}
\end{equation}
with
\begin{equation*}
\tilde{b}^{p,\pm}_{\ell\ell'}=\sum_{p'\neq p}\frac{\tilde{y}^\ell_{p,\mathbf{k}}\tilde{x}^\ell_{p',\mathbf{k}}\tilde{y}^{\ell'}_{p,\mathbf{k}}\tilde{x}^{\ell'}_{p',\mathbf{k}}+\tilde{x}^{\ell}_{p,\mathbf{k}}\tilde{y}^{\ell}_{p',\mathbf{k}}\tilde{x}^{\ell'}_{p,\mathbf{k}}\tilde{y}^{\ell'}_{p',\mathbf{k}}}{2(f^{(0)}_{p,\mathbf{k}}-f^{(0)}_{p',\mathbf{k}})}.\label{b1}
\end{equation*}

Using \eqref{tildefapprox}, we expand the free energy near $\Delta_\ell\simeq0$ and obtain 
\begin{align}
&r'_{\ell\ell'}=\frac{\delta_{\ell\ell'}}{u}-\sum_{p}\int_\mathbf{k}\Big[\frac{\beta}{2}(a^{p,+}_\ell a^{p,+}_{\ell'}+a^{p,-}_\ell a^{p,-}_{\ell'})\nonumber\\
&\times\mathrm{sech}^2\Big(\frac{\beta z_p}{2}\Big)+2(b^{p,+}_{\ell\ell'}+b^{p,-}_{\ell\ell'})\tanh\Big(\frac{\beta z_p}{2}\Big)\Big],
\label{rll''}
\end{align}
where $z_p=f^{(0)}_{p,\mathbf{k}}-\mu$. 

\bibliography{SurNem}

\end{document}